\documentclass[twocolumn, times]{openjournal}

\usepackage{xcolor}
\usepackage{textgreek}
\usepackage[utf8]{inputenc}
\usepackage[english]{babel}

\usepackage{hyperref}
\hypersetup{
    unicode, 
    colorlinks=true,
    linkcolor=linkcolor,
    citecolor=linkcolor,
    filecolor=linkcolor,
    urlcolor=linkcolor,
}
\usepackage{color,colortbl}
\definecolor{linkcolor}{rgb}{0.0,0.3,0.5}
\usepackage{tensind}
\tensordelimiter{?}
\DeclareGraphicsExtensions{.bmp,.png,.jpg,.pdf}
\usepackage{verbatim}
\usepackage[normalem]{ulem}
\usepackage{orcidlink}
\usepackage{soul}
\usepackage{amsmath}

\graphicspath{ {./Figs/} } 
\begin{document}

\title{Viscously Spreading Accretion Disks around Black Holes:  Implications for TDEs, LFBOTs and other Transients}

\author{\vspace{-1.3cm}
Mila Winter-Granic\,\orcidlink{0009-0005-8908-4312}$^{1,\star}$, Eliot Quataert\orcidlink{0000-0001-9185-5044
}$^{1}$}
\email{milawinter@princeton.edu}
\affiliation{$^{1}$Department of Astrophysical Sciences, Princeton University, Princeton, NJ 08544, USA}

\begin{abstract}
We present a simple time-dependent model of viscously spreading accretion disks around black holes (BHs) with masses ranging from $10 -10^8M_\odot$.   We apply the results to observations of late-time emission in tidal disruption events (TDEs) and luminous fast blue optical transients (LFBOTs) such as AT2018cow.  Our model generalizes previous work by incorporating outflows during phases of super-Eddington accretion, non-conservation of mass and angular momentum in TDE circularization, irradiation of the outer disk by the inner accretion flow, and a range of viscous stress models.   We show that many of the late-time plateaus in TDEs can be explained by disks that form with a large spread in angular momentum, due to angular momentum redistribution during circularization.   Viscous spreading on year timescales is not required, although it is also compatible with the data.  The significant range of peak TDE X-ray luminosities is also consistent with a range of disk spreading timescales. The collapse of radiation pressure dominated thin disks to the stable gas-pressure dominated phase underpredicts TDE plateau luminosities by orders of magnitude, strongly favoring thermally stable magnetically dominated disk models.   Irradiation of the outer disk in TDEs due to misalignment of the stellar orbit and BH spin increases plateau luminosities and durations by factors of a few.  Continued study of late-time TDE emission  provides a unique opportunity to constrain the physics of disk formation and circularization, accretion disk warps, angular momentum transport, and other poorly understood aspects of disk physics.   The models developed here can also explain the late-time optical-UV emission in the LFBOT AT2018cow for BH masses of $\sim 10-100 M_\odot$.   The very faint X-ray emission at late-times in AT2018cow is likely due to ongoing X-ray absorption. Our models predict that late-time X-rays at $\sim 10^{39-40} \, {\rm erg \, s^{-1}}$ should eventually be detectable (again) in LFBOTs and that HST-JWST observations of AT2018cow may detect a break in the SED at near-IR-optical wavelengths, providing a powerful probe of the outer accretion disk thermodynamics.

\end{abstract}

\section{Introduction} \label{sec:intro}

 Tidal disruption events (TDEs) are a laboratory to study the nature of accretion disks around compact objects. A TDE occurs when a star passes too close to a massive black hole (BH), such that it is torn apart by the BH's strong tidal forces \citep{hills1975,rees_tidal_1988} producing a bright multiband electromagnetic transient. The early time emission in TDEs is still uncertain theoretically, primarily because of uncertainties in how the highly elliptical stellar debris stream circularizes \citep{hayasaki+2013,dai+2013,shiokawa+2015,bonnerot+2016,metzger_bright_2016,svirski_elliptical_2017,lu_bonnerot2020}.  In addition, the fallback is initially super-Eddington for BHs with $M_\bullet \lesssim 10^7 M_\odot$, which inhibits the cooling needed for the stellar debris to become tightly bound to the BH.   However, at sufficiently late-times the thermal emission in TDEs is likely dominated by an initially compact geometrically thin, optically thick accretion flow that viscously spreads to larger radii over time \citep{cannizzo}. 
 
 Observations indicate that the optical/UV emission from TDEs undergoes a late-time plateau, which has been interpreted as a signature of the emergence of the canonical accretion disk predicted by theoretical models \citep{van_velzen_late-time_2019}.
Recent efforts have fit theoretical models to these late-time observations, as they can provide a unique probe of accretion disk angular momentum transport and the BH mass function in otherwise quiescent galaxies \citep{wen2020,mummery_balbus2020,wen2023,cao_rapidly_2023,mummery_fundamental_2023,mummery_fitting_2025}. 

The physics of spreading disks around BHs is relevant whenever a relatively compact disk is formed quickly, i.e., on timescales less than a viscous time. This includes a wide range of applications beyond those of parabolic TDEs. For example, some theoretical models for luminous fast blue optical transients \citep[LFBOTs;][]{drout+2014,ho+2023} have invoked mergers of a star and a stellar mass BH  \citep{soker_diversity_2019,Metzger2022,Klencki2025,Tsuna2025}. Such models also generically predict late-time emission from a spreading accretion disk formed in the merger.  There is indeed evidence for a slowly fading optical-UV source years after the prototypical LFBOT AT2018cow \citep{sun2022,Chen2023,inkenhaag2023,inkenhaag2025}.

 In this paper we develop a simple model for the dynamics and emission of initially compact accretion disks formed around BHs. The motivation is to easily explore a range of models applicable to BHs of all masses and to incorporate physical ingredients that have not been studied in detail before but that may be important for determining the emission from viscously spreading disks in TDEs, LFBOTs, and other applications.   One aspect of accretion disk physics that we explore in our modeling is that the standard $\alpha$-disk model developed by \cite{shakura&sunyaev} predicts thermal and viscous instabilities in radiation-pressure dominated regimes \citep{lightman_instabilities,ss_instabilites}; these are thought to produce global limit-cycle behavior and large-amplitude variability \citep{honma_unstable,szuszkiewicz_limit-cycle_1998,ohsuga,janiuk2002,shen_evolution_2014,lu2022,piro_late-time_2025}. However, a large majority of X-ray binaries (XRBs) and Active Galactic Nuclei (AGN) lack observational evidence for these instabilities, with the exception of a small set of XRBs that exhibit ``heartbeat oscillations'' with periods that roughly match the Lightman-Eardley prediction \citep[see][]{taam1997}.  One explanation is that strong magnetic fields can stabilize disks against radiation pressure instabilities \citep{begelman_accretion_2007,huang_global_2023,jiang_radiation_2025}. This motivates viscous stress prescriptions that differ significantly from standard gas or radiation pressure dominated models \citep{kaur2023,alush_late-time_2025}.

A second motivation for the present work is that the canonical picture of parabolic TDE disks as having an initial mass of order half the disrupted star's mass and a specific angular momentum equal to that of the disrupted star is not necessarily well motivated.   It is quite possible that angular momentum is not conserved during the process of disk formation in TDEs; e.g., \citet{lu_bonnerot2020} find that redistribution of angular momentum by the self-intersection shock actually leads to accretion disks that are initially retrograde relative to the star's orbit, with significant mass in the disk out to radii larger than (twice) the initial stellar pericenter. In addition, the super-Eddington fallback at early times in TDEs likely leads to optically thick outflows that can unbind a significant fraction of the nominally bound stellar debris \citep{strubbe_optical_2009,ohsuga2011,jiang2014,jiang+2016,huang_global_2023,huang+2024}.  Such super-Eddington outflows will be even more important in stellar merger models for transients such as LFBOTs, since the accretion rates are even higher relative to Eddington than in TDEs by massive BHs.  Another motivation for further exploring the implications of lower initial disk masses is partial tidal disruptions, which may be more common than full disruptions.

Finally, parabolic TDEs are produced by stars on orbits that are uncorrelated with the BH spin.  This presumably leads to accretion disks that are highly inclined with respect to the BH spin.   Assuming that the inner accretion disk at late-times is aligned with the BH spin by the Bardeen-Petterson effect \citep{bardeen_petterson_1975}, irradiation of the outer disk by the inner disk can significantly change the thermal properties of the outer disk that dominates the emission in the optical.   Even in standard AGN there are tentative indications that heating in the optical emitting region is dominated by irradiation rather than local viscous stresses (e.g., via the sizes inferred from microlensing of the disk; \citealt{Dai2010}).

The remainder of this paper is structured as follows. In \S \ref{sec:disk} we lay out the physics of our time-dependent viscously spreading disk models, including effects such as fallback and irradiation. We describe our initial conditions and numerical setup for both the full, radial time-dependent model and the simplified one-zone model in \S\ref{sec:IC}. In \S \ref{sec:applications} we apply these results to TDEs and LFBOTs, and we compare our models with observed data. We conclude and discuss the implications of our results in \S \ref{sec:discussion}.

\section{Disk evolution}
\label{sec:disk}
We base our 1D, time-dependent thin disk model on the standard Shakura-Sunyaev model \citep{shakura&sunyaev}, considering different prescriptions to set the disk viscosity $\nu$. In this model, the gas in the disk moves around the central BH with a Keplerian angular frequency $\Omega=\sqrt{GM_\bullet/R^3}$, where $M_\bullet$ is the BH mass. As the disk evolves, its matter is steadily accreted onto the central BH, such that the disk's surface density $\Sigma(R,t)$ evolves according to the disk diffusion equation

\begin{equation}
    \frac{\partial\Sigma}{\partial t}=\frac{3}{R}\frac{\partial}{\partial R}\left\{R^{1/2}\frac{\partial}{\partial R}\left[\nu\Sigma R^{1/2}\right]\right\}+\dot{\Sigma}_\mathrm{fb}(R,t)-\dot{\Sigma}_\mathrm{out}(R,t),
    \label{eq:diffusion equation}
\end{equation}
where $\nu$ is the effective viscosity, $\dot{\Sigma}_\mathrm{fb}(R,t)$ is a source term that represents mass addition due to fallback and $\dot{\Sigma}_\mathrm{out}(R,t)$ is a sink term that removes mass due to outflows. We adopt the $\alpha$-disk prescription \citep{shakura&sunyaev}, where the disk height $H$ is determined by vertical hydrostatic equilibrium set by the chosen prescription for the disk pressure $P$. The density of the disk is then $\rho=\Sigma/2H$.

The mechanism that provides pressure support within an accretion disk plays a central role in determining its structure and evolution over time. In particular, radiation pressure dominated disks \citep{shakura&sunyaev} are known to be thermally and viscously unstable \citep{Piran1978}, and may as a consequence collapse to the gas pressure dominated branch. Such a collapse, however, can amplify the embedded magnetic field, potentially leading to a magnetically supported state instead \citep[e.g.,][]{begelman_accretion_2007,jiang_radiation_2025}. In order to bracket the uncertainty associated with the disk’s vertical support and angular momentum transport, we consider both possibilities in this work. For the gas pressure dominated case we thus have $\nu_g=\alpha c_s H$ with $H=c_s/\Omega$, while for the magnetized case we will instead use the Alfvén speed such that $\nu_\mathrm{mag}=\alpha v_{A}H$ with $H=v_A/\Omega$. The gas pressure is
\begin{equation}
    P_g = \frac{\rho k_BT_c}{\mu m_p},
\end{equation}
where $T_c = \tau^{1/4}T_\mathrm{eff}$, with $\tau=\Sigma\kappa_e/2$ the optical depth, $\kappa_e=0.34$ cm$^2$ g$^{-1}$ the Thomson scattering opacity and $k_B$ is Boltzmann's constant. In the case of a magnetized disk, we assume that the MRI is suppressed when the Alfvén speed $v_A=\sqrt{P_\mathrm{mag}/\rho}$ exceeds the geometric mean of the Keplerian velocity $v_K=\sqrt{GM_\bullet/R}$ and the gas sound speed $c_{s,g}=\sqrt{P_g/\rho}$ (following \citealt{begelman_accretion_2007}, based on the linear calculations of \citealt{pessah_psaltis_2005}). This limiting velocity $v_A\sim (v_Kc_{s,g})^{1/2}$ can then be adopted as a measure of the characteristic magnetic pressure, which we can express as
\begin{equation}
P_\mathrm{mag}=v_K\rho\sqrt{\frac{k_BT_c}{\mu m_p}}.
\label{eq:Pmag}
\end{equation}
In this limit, $P_\mathrm{mag}\gg P_g$ such that the magnetic pressure dominates the structure of the disk \citep{begelman_accretion_2007}. 
There is support for this stress prescription in the global radiation simulations of \citet{huang_global_2023} (see, in particular their Fig. 8). That being said, the saturation magnetic field strength in MHD accretion disk simulations is also sensitive to the initial field structure (e.g., \citealt{Zhang2026}) so more work is needed to establish the applicability of equation (\ref{eq:Pmag}). Given this uncertainty we will also consider a range of $\alpha$ in what follows as well as present some results for fixed $H/R$ rather than an explicit stress model.


We can express the disk aspect ratio $H/R$ as
\begin{equation}
    \frac{H}{R}=\frac{2P}{GM_\bullet\Sigma}R^2,
    \label{eq:H}
\end{equation}
which allows us to rewrite the viscosity  $\nu=\alpha(H/R)^2\sqrt{GM_\bullet R}$ in terms of the pressure.  If we assume that the disk is in local thermal equilibrium with viscous heating balancing cooling the effective temperature is 
\begin{equation}
    \sigma T_\mathrm{eff}^4=\frac{9}{8}\nu\Sigma\frac{GM_\bullet}{R^3}.
    \label{eq:Teff}
\end{equation}
This estimate of the thermal emission is made under the assumption of a thin disk, which is relevant for the late-time state of disks around higher BH masses. For lower mass BHs, much of the accretion can be super-Eddington and drive winds, which we will consider briefly in section \ref{subsubsec:superedd outflows}.

We assume that the disks are optically thick and emit isotropically, such that the spectral luminosity $L_\nu$ at a given frequency $\nu$ can be approximated by a blackbody distribution $B_\nu$ as
\begin{equation}
    L_\nu = 4\pi^2{\int^{R_{d}}_{r_\mathrm{in}}}B_\nu(T_\mathrm{eff})rdr.
    \label{eq:luminosity}
\end{equation}
This corresponds to the total luminosity emitted by the disk; if we wanted to determine the observed flux instead, there would be an extra factor of $\sin{i}$ set by our viewing angle relative to the plane of the disk.

Combining vertical hydrostatic equilibrium, radiation diffusion to relate $T_c$ and $T_{\rm eff}$, equations \ref{eq:H} and \ref{eq:Teff}, we obtain for a gas pressure supported disk
\begin{equation}
    \nu_g=\left[\frac{9\kappa_e\alpha^4}{16\sigma GM_\bullet}\left(\frac{k_B}{\mu m_p}\right)^4\right]^{1/3}r\Sigma^{2/3},
    \label{eq:nu_g}
\end{equation}
and for a magnetized disk \citep{alush_late-time_2025}
\begin{equation}
\nu_\mathrm{mag}=\left[\frac{9\kappa_e\alpha^8GM_\bullet}{16\sigma}\left(\frac{k_B}{\mu m_p}\right)^4\right]^{1/7}r^{5/7}\Sigma^{2/7},
    \label{eq:nu_mag}
\end{equation}
where $\sigma$ is the Stefan-Boltzmann constant.

Using equations (\ref{eq:nu_g}) and (\ref{eq:nu_mag}) we can obtain an expression for $H/R$ as a function of BH and disk properties as

\begin{equation}
\begin{split}
   \left(\frac{H}{R}\right)_g = 0.004\left(\frac{\alpha}{0.01}\right)^{1/6}\left(\frac{M_\bullet}{10^6M_\odot}\right)^{-5/9}\\
   \left(\frac{M_d}{0.5M_\odot}\right)^{1/3}\left(\frac{R_d}{2r_t}\right)^{-5/12},
   \label{eq:Hgas}
\end{split}
\end{equation}
for a gas pressure supported disk, and

\begin{equation}
\begin{split}
   \left(\frac{H}{R}\right)_\mathrm{mag} = 0.1\left(\frac{\alpha}{0.01}\right)^{1/14}\left(\frac{M_\bullet}{10^6M_\odot}\right)^{-5/21}\\
   \left(\frac{M_d}{0.5M_\odot}\right)^{1/7}\left(\frac{R_d}{2r_t}\right)^{-5/28},
      \label{eq:Hmag}
   \end{split}
\end{equation}
for a magnetized disk, where we have scaled to the standard initial conditions of $M_d=0.5M_\star$ and $R_d=2r_t$ with $r_t$ the tidal radius, assuming a solar mass star.
\subsection{Irradiation}
\label{subsec:irradiation}
The outer regions of a disk may be subject to heating due to irradiation from photons coming from the inner disk in addition to viscous heating.  This is particularly true for parabolic TDEs since for a spinning BH it is a priori likely that the orbital plane of the disrupted star is randomly oriented relative to the spin of the BH.   The thin accretion disk that forms at late-times around the BH can have its angular momentum aligned with that of the BH at small radii, interior to the warp radius \citep{Natarajan1998}
\begin{equation}
r_w \simeq r_g \, 2^{7/3} \, a^{2/3} \, \alpha^{2/3} \, (H/R)^{-4/3}
\label{eq:warp}
\end{equation}
where $a$ is the dimensionless spin of the BH and this expression is only valid for $H/R \lesssim \alpha$. It is useful to compare the warp radius to a fiducial outer disk radius of $2 r_{t}$ corresponding to a circular orbit with the same specific angular momentum as that of the star (though as we have argued in the Introduction, this may not be a good approximation to the disk angular momentum).  This yields
\begin{equation}
\begin{split}
\frac{2 r_t}{r_w} \simeq \frac{0.2}{a^{2/3}} \left(\frac{\alpha}{0.1}\right)^{-2/3}\left(\frac{M_\bullet}{10^6 M_\odot}\right)^{-2/3} \\\left(\frac{H/R}{0.03}\right)^{4/3} \left(\frac{M_\star}{M_\odot}\right)^{1/2},
\label{eq:rd0vsrw}
\end{split}
\end{equation}
where we have assumed a main sequence star with $R_\star\propto M_\star^{0.8}$.
Equation (\ref{eq:rd0vsrw}) shows that the warp radius is typically larger than the fiducial circularization radius of $2 r_t$.  However, the relevant comparison is between $r_w$ and the outer disk radius at the time when $H/R \lesssim \alpha$, which is when Bardeen-Petterson alignment begins to operate.  Since TDE disks will spread significantly prior to reaching $H/R \lesssim \alpha$,  the bulk of the disk will plausibly be exterior to the warp radius at late times, thus spreading in the plane set by the star's initial angular momentum.   Since this plane is highly misaligned with the BH spin, the inner disk can easily irradiate the outer disk, modifying its temperature and emission. 

The arguments above assume that the response of the disk to the warp is in the diffusive regime, valid when $H/R \lesssim \alpha$. Our focus on this regime is motivated by the application to late times in TDE disks when the diffusive regime is most likely to be applicable.    When $H/R \gtrsim \alpha$, however, as is particularly likely to be the case at early times in TDE disks, the disk can instead precess globally in response to the Lens-Thirring torques (e.g., \citealt{Ogilvie1999}).    During the global precession phase the orientation of the disk angular momentum damps viscously and approaches the black hole spin direction, though on a timescale much longer than the global precession timescale \citep{Foucart2014}.  In the following analysis we assume that in the initial thick disk phase the accretion disk in TDEs does not significantly reorient to align with the black hole spin.   \citet{Franchini2016} concluded that significant damping of the disk-BH misalignment is likely during the thick phase. Preliminary calculations that account for viscous spreading suggest that this result may change when viscous spreading of the disk is accounted for because the spreading disk lengthens both the precession and misalignment damping times (Mummery \& Quataert, in prep). However, a detailed treatment of this effect remains beyond the scope of this work. We therefore assume that substantial misalignment can persist throughout the thick disk phase. It is less clear whether the misalignment survives the transition to the thin disk phase as $H/R$ decreases and the viscous spreading slows down. This is, however, plausible given that the response of the disk is also transitioning to the diffusive warp regime.  The assumption of a significant warp at late times is also valid if the early time super-Eddington thick disk phase leads primarily to outflows rather than a coherent disk, as we will argue in \S \ref{subsec:superEdd fallback}: in this case there is no significant circularization until the thin disk phase when the diffusive warp regime likely applies.  In what follows we assume that the warp remains to late times but acknowledge that the early time warp evolution and damping is an interesting and subtle problem that requires additional work.

We define an additional contribution to the local flux due to irradiation as 
\begin{equation}
    q_\mathrm{irr}(r)=f_\mathrm{irr}\frac{L(r_\mathrm{in})}{4\pi r^2}
\end{equation}
for $r>r_w$, where $f_\mathrm{irr}$ is the fraction of the emission from the inner edge of the disk that reaches the outer disk. We estimate this fraction through an irradiation angle $\theta_\mathrm{irr}$ such that $f_\mathrm{irr}\approx\sin{\theta_\mathrm{irr}}$. If $\theta_w$ is the warp angle and $\theta_\star$ is the angle between the orbital plane of the disrupted star and the spin of the BH, then $\theta_\mathrm{irr}=\theta_w-\theta_\star$ with $\tan \theta_w=r\sin\theta_\star/(r\cos\theta_\star-r_w)$ for $r>r_w$.   The heating due to irradiation can in principle be larger or smaller than this fiducial estimate, e.g., if outflows scatter emission to the surface of the outer disk and/or if the outer disk is shadowed by optically thick outflows.   

We note that in the regime where the central temperature of the disk is greater than the effective temperature set by irradiation, only the surface layers of the disk are heated so that its structure and viscous spreading remain unaltered. If the opposite is true, then the disk becomes isothermal which changes its thickness and internal structure, affecting its subsequent evolution. In general for the cases treated in this paper, we are in the former, simpler regime.

\subsection{One-zone model}

\label{subsec:one-zone}

A simple `one-zone' model can be used to approximate the evolution of accretion disks \citep[e.g.][]{metzger_time-dependent_2008,shen_evolution_2014}. The simplicity of this model allows us to easily incorporate additional physics into the disk evolution, such as mass and angular momentum loss from outflows.  At some points in this paper, we will use this simpler model in addition to the spreading disk solutions of equation (\ref{eq:diffusion equation}).

In the one-zone model we focus on the dynamics of the outer part of the disk at radius $R_d$, which we approximate as containing all of the disk's mass $M_d$ and angular momentum $J_d$. Following \cite{metzger_time-dependent_2008}, we set $M_d = A\pi\Sigma R_d^2$ and $J_d=B(GM_\bullet R_d)^{1/2}\pi \Sigma R_d^2$; the constants $A=1.62$ and $B=1.33$ account for the distinction between the total mass of the disk and the mass of the material near $R_d$. We obtain these values by numerically fitting our one-zone model to solutions of equation (\ref{eq:diffusion equation}).

The time evolution of the disk will be determined by the conservation equations. For mass, this gives
\begin{equation}
    \frac{dM_d}{dt}  = -\frac{fM_d}{t_\mathrm{visc}}+\dot{M}_\mathrm{fb},
    \label{eq:Md_dt}
\end{equation}
where $fM_d/t_\mathrm{visc}=\dot{M}_\mathrm{in}+\dot{M}_\mathrm{out}$ is the outer disk inflow rate split into  accretion at small radii plus mass loss due to winds, with $f\simeq1.6$ set to match the calibrations with $A$ and $B$, and $\dot{M}_\mathrm{fb}$ is the fallback rate (see sections \ref{subsubsec:superedd outflows} and \ref{sec:IC}, respectively, for our models of these processes).

Conservation of angular momentum gives
\begin{equation}
    \frac{dJ_d}{dt}=\dot{J}_\mathrm{fb}-\dot{J}_\mathrm{out},
    \label{eq:Jd_dt}
\end{equation}
where $\dot J_{\rm out}$ is the angular momentum lost to winds, $\dot{J}_\mathrm{fb}=\dot{M}_\mathrm{fb}(GM_\bullet r_c)^{1/2}$, and  $r_c$ is the circularization radius. The angular momentum of the disk is conserved in the simplest scenario, which includes only viscous spreading and no fallback or outflows. We note that strictly speaking, an extra sink term should be taken into account due to loss into the plunging region. This term in general will be small, except for disks with circularization radii close to the last stable circular orbit (e.g., high $\beta$ or high black hole masses).

\begin{figure}
\centering
\includegraphics[width=\linewidth]{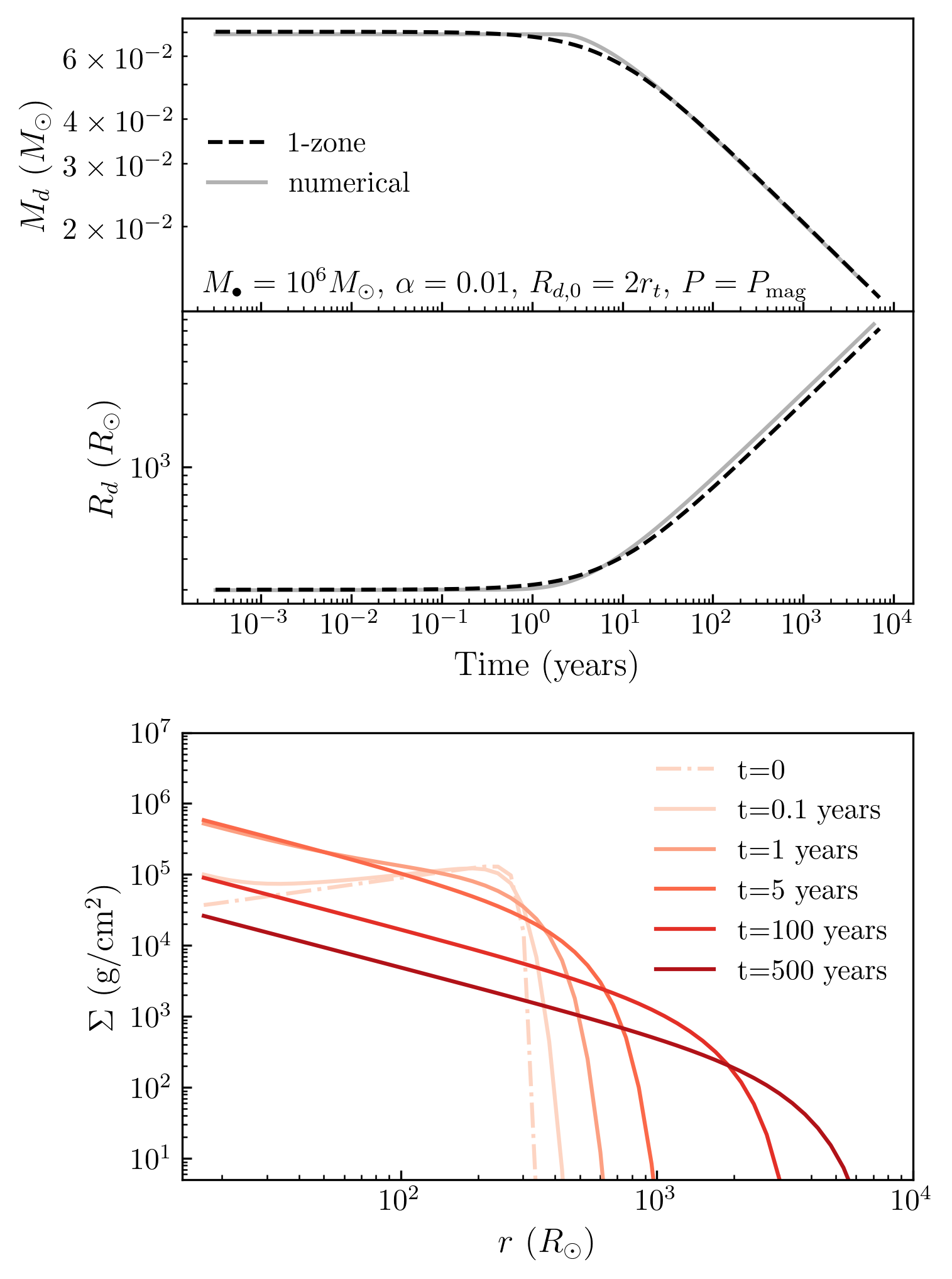}
\caption{Full radial, time-dependent evolution of a magnetized disk. Top two panels show the evolution of the disk mass $M_d$ and disk radius $R_d$ respectively, where we have included results obtained using the 1-zone model for comparison purposes. The initial disk mass and radius are set to $M_{d,0}=0.07M_\odot$ (see equation (\ref{eq:Md0}) in \S\ref{subsubsec:superedd outflows}) and $R_{d,0}=2r_t$. Bottom panel shows the temporal evolution of the surface density profile up to $t=500$ years, with the initial state shown with a dot-dashed line.}
\label{fig:evolution}
\end{figure}

In Figure \ref{fig:evolution} we show an example of the full radial, time-dependent evolution of a magnetized disk with $M_\bullet=10^6M_\odot$ and $\alpha=0.01$, along with analogous results obtained through the one-zone approximation. In this example we initialize the disk with a mass of $M_{d,0}=0.07M_\odot$ and radius of $R_{d,0}=2r_t$.
Both models agree nicely, confirming the validity of the 1-zone approximation. We note that our magnetized disk model produces results consistent with those of \cite{alush_late-time_2025}.

\subsection{Super-Eddington outflows}
\label{subsubsec:superedd outflows}
During the stages when a disk is accreting at super-Eddington rates, taking into account mass and angular momentum loss via outflows becomes critical. The one-zone model is particularly useful for this purpose due to its simplicity. 

For lower mass BHs, super-Eddington accretion can extend all the way out to the outer edge of the disk at early times, which requires a different treatment of the disk's evolution as it will be geometrically thick and radiatively inefficient. We will assume that only a fraction 
$\sim (r_\mathrm{in}/R_d)^p$ of the available material is accreted onto the central BH, and that the remaining material is lost to an outflow. Here we follow \cite{blandford} and use $p$ to quantify the strength of outflows. In the absence of outflows $p=0$.  In this work we use $p=0.5$ as motivated by simulations by \cite{guo_cyclic_2025}. In order to include mass loss due to super-Eddington winds, we employ a one-zone spreading ring approximation with fixed scale height $H/R=1/3$. The outflow is parametrized using 
\begin{equation}
    \dot{M}_\mathrm{out}=\left[1-\left(\frac{r_\mathrm{in}}{R_d}\right)^p\right]\frac{fM_d}{t_\mathrm{visc}},
    \label{eq:Mdot_out}
\end{equation}
and the accretion rate onto the BH is then given by
\begin{equation}
    \dot{M}_\mathrm{in}=\left (\frac{r_\mathrm{in}}{R_d}\right)^p\frac{fM_d}{t_\mathrm{visc}}.
    \label{eq:Mdot_in}
\end{equation}
Here
\begin{equation}
   t_\mathrm{visc}=\frac{R_d^2}{\nu}.
\end{equation}
The initial viscous time, $t_\mathrm{visc,0}$, essentially sets the amount of time it takes for material on the outer edge of the disk to start being accreted onto the central BH and for the disk to begin spreading outwards (although typically the disk radius doubles on a timescale that is a few times shorter than $t_\mathrm{visc,0}$).

The angular momentum loss rate from the disk due to the outflows is then
\begin{equation}
    \dot{J}_\mathrm{out}=C\dot{M}_\mathrm{out}(GM_\bullet R_d)^{1/2} \label{eq:Jdot_out}
\end{equation}
where $C$ is a constant that depends on the torque exerted by the outflowing mass on the remaining disk. Simulations from \cite{guo_cyclic_2025} suggest $p\approx C\approx0.5$ for this regime. This gives a disk mass that evolves as
\begin{equation}
    M_d\simeq M_{d,0}\left[1+2.4\left(\frac{t}{t_\mathrm{visc,0}}\right)\right]^{-2/3},
\end{equation}
and a disk radius
\begin{equation}
    R_d\simeq R_{d,0}\left[1+2.4\left(\frac{t}{t_\mathrm{visc,0}}\right)\right]^{2/3},
\end{equation}
where the exponent and 2.4 pre-factor come from assuming $C=0.5$ and $f\simeq1.6$ which is set to match the calibrations with the one-zone model parameters $A$ and $B$ as mentioned in section \ref{subsec:one-zone}.
At large radii, the inflow rate is given by 
\begin{equation}
    \dot{M}(R_d)=-\frac{dM_d}{dt}=\frac{1.6M_{d,0}}{t_\mathrm{visc,0}}\left[1+2.4\left(\frac{t}{t_\mathrm{visc,0}}\right)\right]^{-5/3}.
\end{equation}
The outer disk becomes sub-Eddington when $GM_\bullet\dot{M}(R_d)/R_d=L_\mathrm{Edd}$, after which it collapses into a thin and radiatively efficient disk, with its height and spreading rate set by the assumed pressure support, as discussed in the first part of this Section.

\section{Initial Conditions}
\label{sec:IC}

In this work we will focus  on disks that are formed due to stars being tidally disrupted by a BH. We consider both stars on bound circular orbits (e.g., stellar mergers or stars inspiraling onto massive BHs by gravitational wave radiation as in \citealt{linial_tidal_2024}) and parabolic TDEs.  

\subsection{Bound Stars on Circular Orbits}

The initial conditions for tidal disruption of a bound star on a circular orbit are particularly straightforward with $M_{d,0}=M_\star$ and $R_{d,0}=r_t$ where 
\begin{equation}
    r_t = R_\star\left(\frac{M_\bullet}{M_\star}\right)^{1/3}.
\end{equation}
We focus on the one-zone models for our models of stellar mergers and TDEs on circular orbits to easily incorporate outflows and explore a large parameter space of models.  This amounts to integrating equations (\ref{eq:Md_dt}) and (\ref{eq:Jd_dt}) over time.   

\subsection{Parabolic TDEs}

For stars on roughly parabolic orbits disrupted by a massive BH, we assume that the disk forms with a characteristic outer radius of 
\begin{equation}
    R_{d,0}=\frac{2}{\beta}r_tf_J,
\end{equation}
where $\beta$ is the ratio of the tidal radius $r_t$ to the pericenter radius, and we take $\beta=1$.  The factor $f_J$ quantifies the extent to which angular momentum is conserved during disk formation and in principle can be either $> 1$  or $< 1$.   As we discussed in the Introduction, some simulations favor $f_J > 1$ due to angular momentum redistribution by circularization shocks \citep{lu_bonnerot2020}.  We will consider both $f_J = 1$ and $f_J > 1$ in what follows.

\subsubsection{Super-Eddington Fallback}
\label{subsec:superEdd fallback}
In a parabolic TDE, the loosely bound stellar debris falls back towards the BH at a rate
\begin{equation}
    \dot{M}_\mathrm{fb}=\frac{M_\star}{3t_\mathrm{fb}}\left(\frac{t}{t_\mathrm{fb}}\right)^{-5/3}
    \label{eq:Mdot fb}
\end{equation}
for $t\gtrsim t_\mathrm{fb}$, where \citep{Bandopadhyay2024}
\begin{equation}
     t_\mathrm{fb}\simeq 30\text{ d}\left(\frac{M_\bullet}{10^6M_\odot}\right)^{1/2}.
\end{equation}
Fallback is initially super-Eddington for BHs with masses $M_\bullet \lesssim {\rm few} \times 10^7 M_\odot$.

Fallback only becomes sub-Eddington at a time
\begin{equation}
    t_\mathrm{Edd}\simeq1.9\text{ yr}\left(\frac{M_\bullet}{10^6M_\odot}\right)^{-2/5}\left(\frac{M_\star}{M_\odot}\right)^{3/5}.
    \label{eq:tedd_fb}
\end{equation}
We strongly suspect that the initial super-Eddington fallback leads to much of the stellar debris being unbound (as is found in, e.g., the self-intersection shock simulations of \citealt{jiang+2016,huang+2024} and global 3D MHD simulations of \citealt{jiang2014, huang_global_2023}).   This will lead to initial disk masses  $M_{d,0}\leq M_\star/2$.  To estimate the magnitude of the bound mass we consider a toy model in which we solve equation (\ref{eq:diffusion equation}) with fallback as a mass source and assume that 
at super-Eddington rates the disk has $H/R=1/3$ but only accumulates mass at a slower rate of $f_\mathrm{fb}\dot{M}_\mathrm{fb}$ with $f_{\rm fb} < 1$ because of outflows during super-Eddington circularization.\footnote{Existing simulations of super-Eddington accretion that motivate equations (\ref{eq:Mdot_out}) and (\ref{eq:Jdot_out}) with $p \sim C \sim 1/2$ are for disks that start tightly bound to the BH.  The case of TDE circularization and disk formation during the super-Eddington fallback phase is significantly more complex and the exact outflow of mass and angular momentum is not as well understood.  This is why we explore a parameterized toy model here rather than use equations (\ref{eq:Mdot_out}) and (\ref{eq:Jdot_out}) for super-Eddington fallback in parabolic TDEs.}   For $t > t_{\rm Edd}$, the disk becomes thin and radiatively efficient, and continues accumulating material at the standard fallback rate, i.e., with $f_{\rm fb} = 1$.

We find that as long as $f_{\rm fb} \lesssim 0.1$ during the super-Eddington phase the exact bound disk mass at late-times is insensitive to $f_{\rm fb}$ and is 
\begin{equation}
    M_\mathrm{d,0}\sim 0.07 M_\odot\left(\frac{M_\star}{M_\odot}\right)^{3/5}\left(\frac{M_\bullet}{10^6M_\odot}\right)^{3/5},
    \label{eq:Md0}
\end{equation}
which one can obtain by either analytically estimating the mass available when the fallback rate falls below Eddington, or performing a numerical fit to our toy model solution to equation (\ref{eq:diffusion equation}).
We find similar expressions for either the magnetized disk or gas pressure dominated models so given the order of magnitude nature of this estimate we do not distinguish between the two in what follows.   The disk evolution during the sub-Eddington fallback phase is the same whether we start with a disk mass of equation (\ref{eq:Md0})  or integrate equation (\ref{eq:diffusion equation}) with suppressed fallback.   For simplicity we thus solve our parabolic TDE models using equation (\ref{eq:Md0}) as our initial disk mass and do not explicitly include fallback in the calculations.  We will also compare our results to the standard assumption of $M_{d,0} = 0.5 M_\star$.
We note that for parabolic TDEs by intermediate mass BHs,   $t_\mathrm{Edd}$ becomes of order a decade or longer.   Predictions for the emission associated with such events on observable timescales will thus be particularly sensitive to the uncertain physics of super-Eddington fallback and disk formation.

\subsubsection{Numerical setup}

In our models of parabolic TDEs, we begin with an initial power-law surface density profile set by
\begin{equation}
    \Sigma(r,0) = \Sigma_0\left(\frac{r}{R_{d,0}}\right)^{-\gamma}
    \label{eq:sigma0}
\end{equation}
for $r <R_{d,0}$, where we have arbitrarily taken $\gamma=-0.5$, as the disk evolution shows practically no dependence on this parameter. We define $\Sigma_0$ by normalizing such that $\int_{r_\mathrm{in}}^{R_{d,0}} \Sigma(r,0) 2\pi rdr= M_{d,0}$, and then solve equation (\ref{eq:diffusion equation}) using the explicit Euler method.

\section{Applications}
\label{sec:applications}

In this section we consider the physics of spreading disks and apply it to a range of BH masses, putting it in the context of observed transients such as TDEs and LFBOTs.

\subsection{TDEs}
\label{subsec:TDEs}

Observed TDE BH masses are typically $\sim 10^6-10^8M_\odot$, though TDEs around lower mass BHs are of considerable interest as well as a way to probe intermediate mass BHs.   

Here we focus on parabolic orbits where only a fraction of the disrupted star's mass likely remains bound to the central BH and forms a disk (see section \ref{subsec:superEdd fallback}).

\subsubsection{Non-Spreading Disks}
\label{subsec:non-spread}

\begin{figure}
\centering
\includegraphics[width=\linewidth]{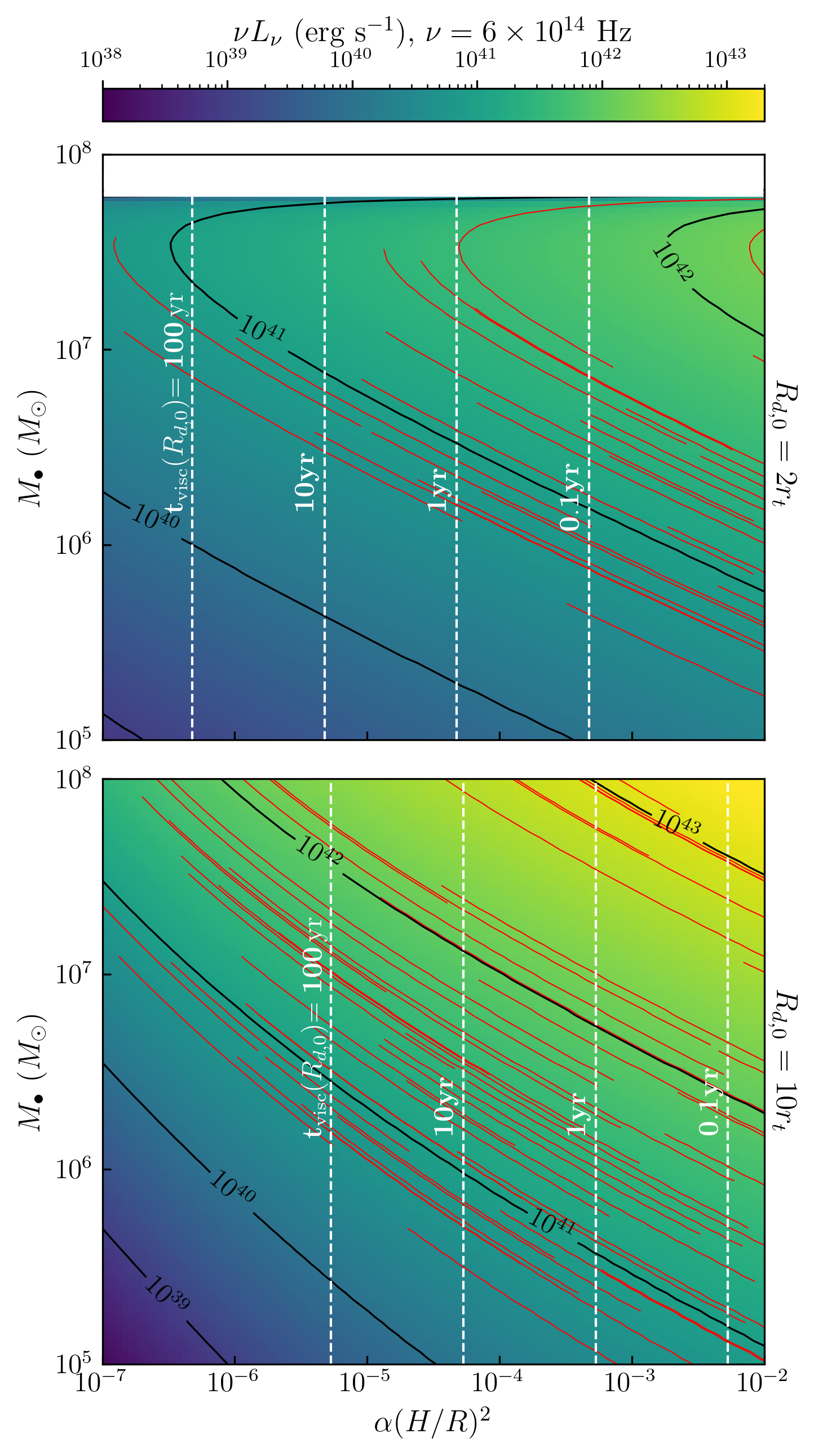}
\caption{Initial disk luminosities prior to viscous spreading with $R_{d,0}=2r_t$ and $R_{d,0}=10r_t$, for different $M_\bullet$ and $\alpha(H/R)^2$ combinations. All models have disk mass set by equation (\ref{eq:Md0}). For reference, a fiducial model with $M_\bullet=10^6M_\odot$ and $\alpha=0.01$ gives $(H/R)_g\sim0.003$ and $(H/R)_\mathrm{mag}\sim0.06$ for a gas and magnetic pressure supported disk, respectively. Black contours show a few reference luminosities, while red contours represent observed values of plateau luminosities for 49 observed TDEs from \cite{mummery_fundamental_2023}. We only include those in the sample that have a measured plateau luminosity, and have truncated the BH masses they cover in the plot such that they only extend over a range of plausible masses suggested by either $M_\bullet-\sigma$ or $M_\bullet-M_\mathrm{gal}$ relations. We include in white dashed lines the viscous timescale $t_\mathrm{visc}(R_{d,0})$. Many of the observed late-time TDE luminosities fall within the parameter space of non-spreading disks with long viscous timescales, which suggests that they can be explained by disks that form with a large spread of angular momentum.}
\label{fig:constant_lum}
\end{figure}

In this section we explore the possibility that disks in TDEs form with an initially large spread in angular momentum.
In this case the lack of significant late-time evolution would in part reflect the long viscous time of the outer disk.  This possibility is motivated by the significant redistribution of angular momentum found in some simulations of TDE circularization and disk formation \citep{lu_bonnerot2020}.

By taking the initial conditions presented in \S\ref{sec:IC}, we can estimate what the luminosity of a disk with an initial radius of $R_{d,0}=2r_t$ or $R_{d,0}=10r_t$ would be; here we are assuming that the disks are in thermal equilibrium but not viscous equilibrium. We present these luminosities for a range of BH masses and $\alpha(H/R)^2$ values in Figure \ref{fig:constant_lum}, covering reasonable values that would be given by both gas pressure and magnetic pressure supported disks (see eqs \ref{eq:Hgas} and \ref{eq:Hmag}, respectively). We take the initial mass of the disks to be given by equation (\ref{eq:Md0}).

The red contours in this plot indicate observed UV/optical plateau luminosities $\nu L_\nu$ (measured in the rest-frame $g$ band) from \cite{mummery_fundamental_2023}, showing that a wide range of these values can be explained through our model without necessarily invoking viscous spreading. \citet{mummery_fundamental_2023}  measure the plateau luminosities by fitting the full light curve using a two component model:  the first is a Gaussian-rise, exponential decay model (for the optical peak) while the second has a constant flux (the plateau). They assume blackbody spectra, and then fit the light curve model to all  optical/UV photometry simultaneously. 

In Figure \ref{fig:constant_lum} these curves have been truncated such that they only cover a range of plausible masses for each TDE, suggested by either the $M_\bullet-\sigma$ relation or the galactic mass scaling relation \citep{greene2020} when velocity dispersion measurements for the host were not available (values of $\sigma$ and $M_\mathrm{gal}$ are reported in \citealt{mummery_fundamental_2023}). The white dashed lines indicate the viscous timescales for these models, which must be on the order of at least a few years in order for the non-spreading picture to be consistent. This timescale is given by
\begin{equation}
\begin{split}
    t_\mathrm{visc}(R_{d,0})=1.4\text{ yr }\left(\frac{\alpha(H/R)^2}{10^{-4}}\right)^{-1}
    \left(\frac{R_{d,0}}{2r_t}\right)^{3/2}\left(\frac{M_\star}{M_\odot}\right)^{0.7}.
    \label{eq:t_visc}
\end{split}
\end{equation}
We find from our time-evolving models that the disk radius initially doubles on a timescale of $\sim t_\mathrm{visc}/3$, and we have therefore included this factor of $1/3$ in the values presented in Figure \ref{fig:constant_lum}. From equation (\ref{eq:t_visc}) we can obtain an approximate expression for the luminosity on the Rayleigh-Jeans tail of
\begin{equation}
\begin{split}
    \nu L_\nu\sim 5 \times10^{40}&\text{ erg s}^{-1}\left(\frac{\nu}{6\times10^{14}\text{ Hz}}\right)^3 \left(\frac{M_\bullet}{10^6M_\odot}\right)^{2/3}\\
    &\left(\frac{M_d}{0.07M_\odot}\right)^{1/4}\left(\frac{R_{d,0}}{2r_t}\right)^{5/4}\left(\frac{t_\mathrm{visc}}{1\mathrm{yr}}\right)^{-1/4}.
    \end{split}
    \label{eq:Lnospread}
\end{equation}
We note that for this approximation we have assumed that the emission lies on the Rayleigh-Jeans tail, which is not necessarily the case at larger values of $R_{d,0}$. This leads to a slight over-estimate of the luminosity, as compared to the more accurate calculations presented in Figure \ref{fig:constant_lum}. In Figure \ref{fig:constant_lum} we have also assumed (arbitrarily) that the initial surface density profile  follows a power-law with exponent $\gamma=-0.5$ (see equation \ref{eq:sigma0}).  This is justified for spreading disks as the late-time emission loses memory of the initial shape of $\Sigma(r)$.  At early times this is not true, but we find that as long as most of the initial disk mass is concentrated at large radii then the estimated luminosity does not change by more than order unity for different choices of the initial surface density distribution. 

Figure \ref{fig:constant_lum} and equation (\ref{eq:Lnospread}) show that there is a range of parameters that can explain many of the observed late-time TDE sources through a  disk that is viscously heated but not necessarily spreading significantly -- the brightest sources near $\nu L_\nu\sim10^{43}$ erg s$^{-1}$ only appear in the bottom panel, which implies they require radially dispersed initial conditions and thus limited viscous spreading so that the disk can maintain a high luminosity for long enough.  Equation (\ref{eq:Lnospread}) highlights part of the reason for this: the dependence of the RJ tail emission on $t_\mathrm{visc}$ is relatively weak so that  disks with long viscous timescales can still produce luminosities consistent with those observed at late-times in TDEs, particularly if the disk forms with a wider spread of angular momentum than traditionally assumed (larger $R_d$).   

We conclude then that although we agree with previous work arguing that viscously spreading disks are a good model for optical/UV plateaus in TDEs, the early time luminosities in many cases do not explicitly require spreading on the timescale of the observations.   

A similar conclusion was reached by \cite{wen2023}, who found that observations of the TDE ASASSN-14li are consistent with a barely spreading viscous disk that still has a radius of $\sim2r_t$ at late-times. 

\subsubsection{Viscously Spreading Disks}

\begin{figure}
\centering
\includegraphics[width=\linewidth]{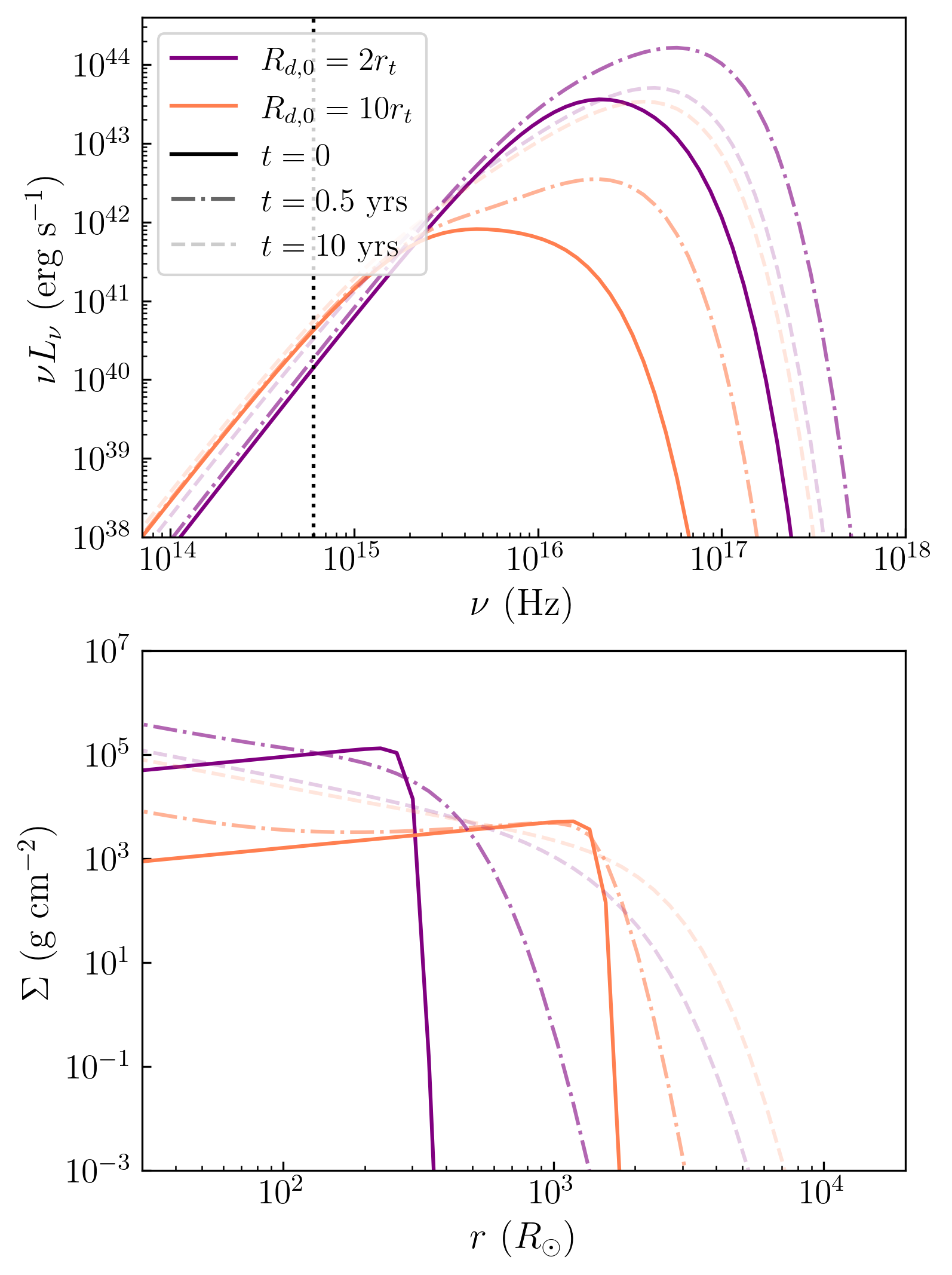}
\caption{SEDs (top panel) and surface density profiles (bottom panel) at different times for a model with $M_\bullet=10^6M_\odot$, $M_{d,0}=0.07M_\odot$ and $\alpha(H/R)^2=10^{-5}$, with initial radii $R_{d,0}=2r_t$ (purple) and $R_{d,0}=10r_t$ (orange). The vertical dotted line marks $\nu=6\times10^{14}$Hz for reference. For different initial disk radii, the initial SEDs (solid lines) differ considerably in where the Rayleigh-Jeans break occurs, which can be used to determine the initial structure of the disk and identify whether it is in fact viscously spreading. At later times, once the disk has spread significantly, the SEDs and surface density profiles are relatively independent of the initial profile.}
\label{fig:sigmas_seds}
\end{figure}

\begin{figure}
\centering
\includegraphics[width=\linewidth]{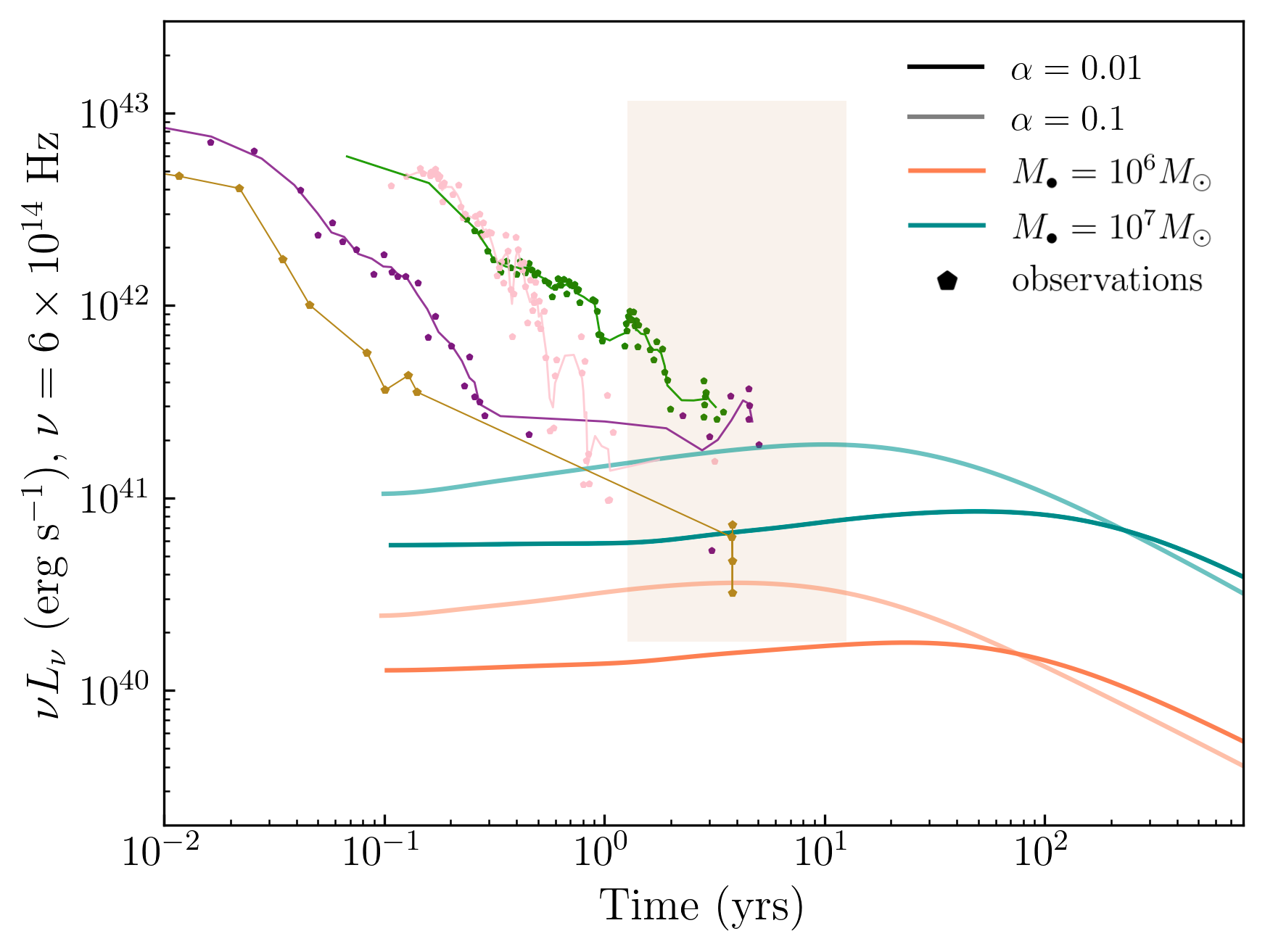}
\caption{Optical light curves for different TDE models with initial disk mass set by equation (\ref{eq:Md0}), and fiducial initial radius of $R_{d,0}=2r_t$. Observed data in the g-band for the events AT2019qiz, AT2020wey, AT2020ocn and AT2021ehb are plotted for comparison purposes. The shaded region covers the range of inferred luminosity plateau values obtained by \cite{mummery_fundamental_2023} over the times the observations were made ($\sim 1-10$yr), which overlaps with the plateaus for our more luminous models. }
\label{fig:TDE_Lnu}
\end{figure}

\begin{figure*}
\centering
\includegraphics[width=\linewidth]{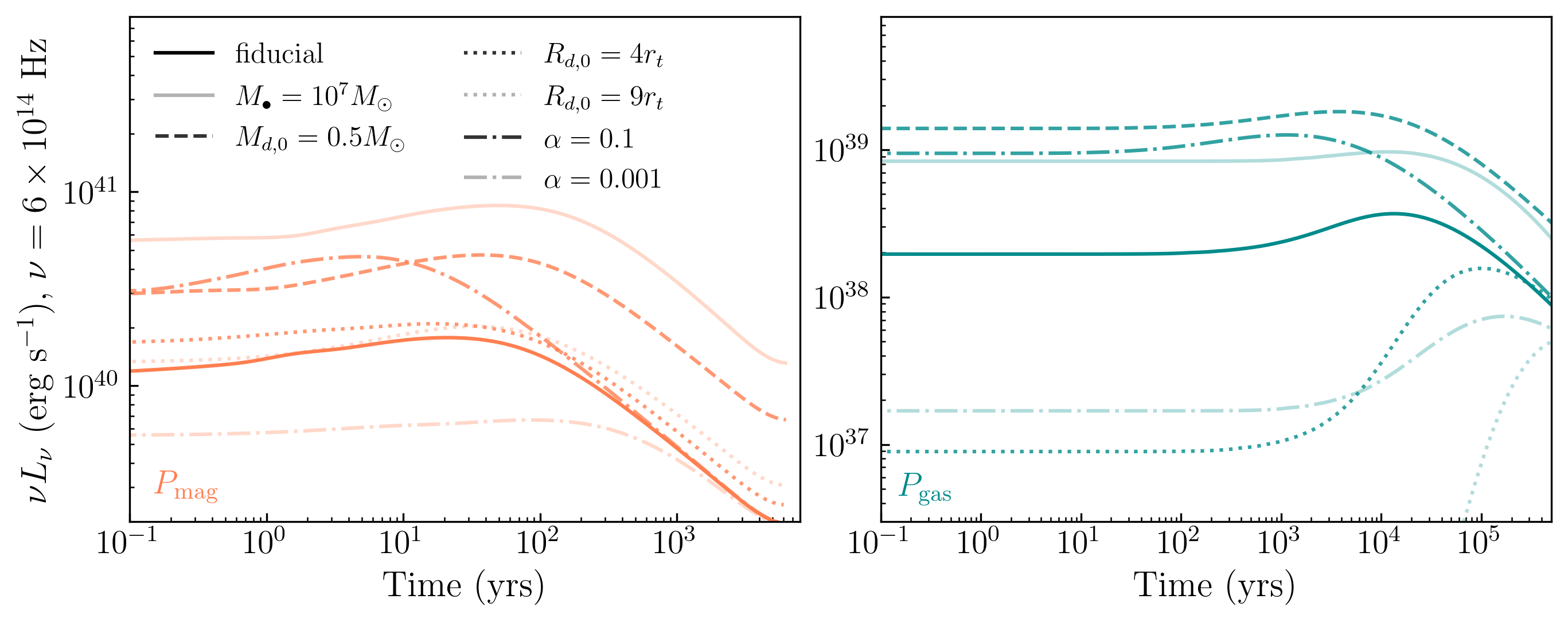}
\caption{Optical light curves for different disk models varying BH mass, initial disk mass, initial disk radius and $\alpha$. The left panel shows models for magnetized disk, while the right panel shows gas pressure supported disks. The fiducial model represents a disk with $M_\bullet=10^6M_\odot$, $R_{d,0}=2r_t$ and $\alpha=0.01$. All models have an initial disk mass calculated according to equation (\ref{eq:Md0}) except when $M_{d,0}$ is specified. Magnetized disks produce overall higher luminosities than gas pressure supported disks, which can be further increased by having larger $M_\bullet$, $\alpha$, $M_{d,0}$, or by increasing $R_{d,0}$ as long as the optical-UV emission is still the Rayleigh-Jeans tail i.e. $k_BT_{\rm eff}(R_{d,0})\sim h\nu$.  Collapse of an unstable radiation pressure supported disk to the gas pressure supported branch produces optical luminosities much fainter than observed (right panel), favoring magnetically supported disk models.} 
\label{fig:TDE_Lnu_comparisons}
\end{figure*}

Let us now consider disks that spread over time. For the range of BH masses we consider in this section, we can plausibly ignore mass loss from the outer edge of the disk as accretion will be sub-Eddington once the disk forms at later times; we thus perform full radial, time-dependent integrations of equation (\ref{eq:diffusion equation}). 

In Figure \ref{fig:sigmas_seds} we show the SEDs and surface density profiles at different times for a  model with $M_\bullet=10^6M_\odot$, $M_{d,0}=0.07M_\odot$ and constant $\alpha(H/R)^2=10^{-5}$, for two different initial disk radii $R_{d,0}$. The solid lines represent the initial state of the disk, which would have luminosities as shown in Figure \ref{fig:constant_lum} and estimated in equation (\ref{eq:Lnospread}). We note that the solid (initial) and dash-dotted ($t = 0.5$ yr) SEDs differ considerably in where the Rayleigh-Jeans break occurs and in the early-time UV and X-ray flux.  Early time observations, particularly in the UV and X-ray, can thus help constrain the initial spread in angular momentum with which the disk forms (we discuss this in more detail below). After a sufficient amount of time, however, the $R_{d,0}=2r_t$ model spreads enough that its surface density profile and hence its SED resembles that of the $R_{d,0}=10r_t$ model quite closely. 

In a more realistic scenario choosing a constant $\alpha(H/R)^2$ is likely inaccurate as the disk aspect ratio $H/R$ is likely to evolve over time. We will now consider disks with physically motivated (albeit still quite uncertain) viscosity prescriptions set by both gas and magnetic pressure through equations (\ref{eq:nu_g}) and (\ref{eq:nu_mag}).

We present some examples of light curves obtained for magnetized disks in Figure \ref{fig:TDE_Lnu}, plotted along with observed luminosity curves for a few TDEs. The shaded region covers the range of inferred luminosity plateau values obtained by \cite{mummery_fundamental_2023} over the times the observations were made ($\sim 1-10$ yrs). As we are focusing on understanding the late-time emission of these events, the early light curves for our models should not be considered when comparing to data. However after $t\gtrsim t_{\rm Edd} \sim $ yr (for BHs with $M_\bullet \gtrsim 10^6 M_\odot$) we assume that the disk has circularized such that our model is reasonably applicable.

It is instructive to analytically quantify how the disk emission on the RJ tail depends on the properties of the TDE and the viscous model. Using $B_\nu\sim T_\mathrm{eff}$ and that most of the mass is at large radii such that $\Sigma\sim M_d/r^2$. We find from equation (\ref{eq:luminosity}) (assuming $R_d\gg r_\mathrm{in}$) that 
\begin{equation}
    L_\nu\propto
    \begin{cases}    M_\bullet^{7/18}\alpha^{1/3}M_d^{5/12}\left(\frac{R_d}{r_t}\right)^{2/3}, & P=P_g\\ M_\bullet^{23/42}\alpha^{2/7}M_d^{9/28}\left(\frac{R_d}{r_t}\right)^{11/14} , &P=P_\mathrm{mag}
    \end{cases}.
\label{eq:Lscalings}
\end{equation}
These scalings highlight the strong dependence of the RJ luminosity on BH mass (as emphasized by \citealt{mummery_fundamental_2023})
and that the luminosity is also sensitive to the initial disk mass and angular momentum.   We plot different models that show this dependence of the disk optical-UV emission for both pressure prescriptions in Figure \ref{fig:TDE_Lnu_comparisons}. We explore a range of initial $M_{d,0}$ and $R_{d,0}$, where in the fiducial models we take the initial disk mass we get from taking into consideration super-Eddington outflows from equation (\ref{eq:Md0}). The left panel of Figure \ref{fig:TDE_Lnu_comparisons} shows that the disk mass, initial radius, and viscosity all impact the early-time luminosity, but their effects are  smaller than the difference between our fiducial $M_\bullet = 10^{6-7} M_\odot$ models.    That being said, by far the biggest difference in the models is  produced by comparing our magnetic pressure vs gas pressure dominated models (left vs. right panels).   This reflects the large difference in $H/R$ and thus viscous heating and emission between these models.    The late-time luminosities in TDEs, if indeed associated with a quasi-steady disk, are incompatible with the disk collapsing to the gas pressure dominated branch as a result of thermal/viscous instabilities. Even the dimmest detected plateau still has a luminosity of $\sim7\times 10^{40}$ erg s$^{-1}$ \citep[see][Fig. 8]{mummery_fundamental_2023}, which is far too bright to be explained by the dimmer luminosities we predict for gas pressure supported disks.   The late-time luminosities are, however, reasonably consistent with the magnetically dominated models considered here.

Figure \ref{fig:TDE_Lnu_comparisons} shows that there is an increase in the early-time luminosity for higher $R_{d,0}$ in the magnetized viscosity model, but this increase is not monotonic. If we consider $R_{d,0}=9r_t$ we see that the luminosity is actually a bit lower than the $R_{d,0}=4r_t$ model, suggesting that for a given model there is a value of $R_d$ that will maximize the early optical/UV luminosity of the disk. This occurs when $k_BT_\mathrm{eff}(R_d)\sim h\nu$ i.e. while the optical/UV remains on the Rayleigh-Jeans tail. We define this radius as
\begin{equation}
\begin{split}
\frac{r_{L_\nu,\mathrm{max}}}{r_t}&\approx10\left(\frac{\alpha}{0.01}\right)^{4/17}\left(\frac{M_\bullet}{10^6M_\odot}\right)^{-5/51}\\
&\times\left(\frac{M_d}{0.07M_\odot}\right)^{9/34}\left(\frac{\nu}{6\times10^{14}\text{Hz}}\right)^{-14/17}.
\end{split}
\label{eq:rLmaxmag}
\end{equation}
for the magnetized disk case, and

\begin{equation}
\begin{split}
\frac{r_{L_\nu,\mathrm{max}}}{r_t}&\approx1\left(\frac{\alpha}{0.01}\right)^{1/4}\left(\frac{M_\bullet}{10^6M_\odot}\right)^{-5/24}\\
&\times\left(\frac{M_d}{0.07M_\odot}\right)^{5/16}\left(\frac{\nu}{6\times10^{14}\text{Hz}}\right)^{-3/4}.
\end{split}
\end{equation}
for the gas pressure supported case. For the latter case, the luminosity is maximized when $R_{d,0}\sim r_t$, meaning that basically any increase in initial disk radius will result in lower luminosities, as is clearly shown in the right panel of Fig. \ref{fig:TDE_Lnu_comparisons}.

Given the conclusion in \S \ref{subsec:non-spread} that both viscously spreading disks and non-spreading disks with larger $R_d$ can produce the late-time optical-UV luminosities observed in TDEs, there is a potential degeneracy in the BH and disk parameters one would infer from observations, depending on which of these scenarios is correct. However these models predict  differences that can in principle be used to discriminate between them observationally. The key discriminant is in the  evolution of the SED, particularly in the UV and X-rays (see also, e.g., \citealt{mummery_balbus2020}).  This is highlighted in Figure \ref{fig:sigmas_seds}, where the model with larger $R_{d,0}$ has both lower early time far UV-X-ray luminosities and a lower peak far UV-X-ray luminosity over all.
There is a significant amount of X-ray data for a number of TDEs, particularly at early times.  As we now discuss, this data can help further distinguish between the models presented in Figure \ref{fig:sigmas_seds}.

\cite{guolo2024}  presents a sample of 17 optically selected, X-ray detected TDEs for which they provide peak X-ray luminosities in the 0.3-10 keV range, along with estimates for their BH masses from galaxy-BH mass correlations.  A simple but accurate analytical estimate of the peak thermal disk X-ray luminosity can be made by noting that this peak coincides with the peak in the accretion rate onto the central BH, which is given roughly by $\dot{M}_\mathrm{peak}\sim M_{d,0}/t_\mathrm{visc,0}$, where the viscous time is evaluated at the radius with most of the initial disk mass. The peak X-ray luminosity is then given by the steady state disk prediction given $M_\bullet$ and $\dot{M}_\mathrm{peak}$, where the latter is
\begin{equation}
\begin{split}
    \dot{M}_\mathrm{peak}\sim \frac{M_{d,0}}{t_{{\rm visc},0}} & \sim 0.05M_\odot \text{ yr}^{-1}\left(\frac{M_{d,0}}{0.07
    M_\odot}\right)\\
    &\left(\frac{R_{d,0}}{2r_t}\right)^{-3/2}\left(\frac{\alpha(H/R)^2}{10^{-4}}\right).
    \label{eq:Mdot_peak}
\end{split}
\end{equation}

\begin{figure}
\centering
\includegraphics[width=\linewidth]{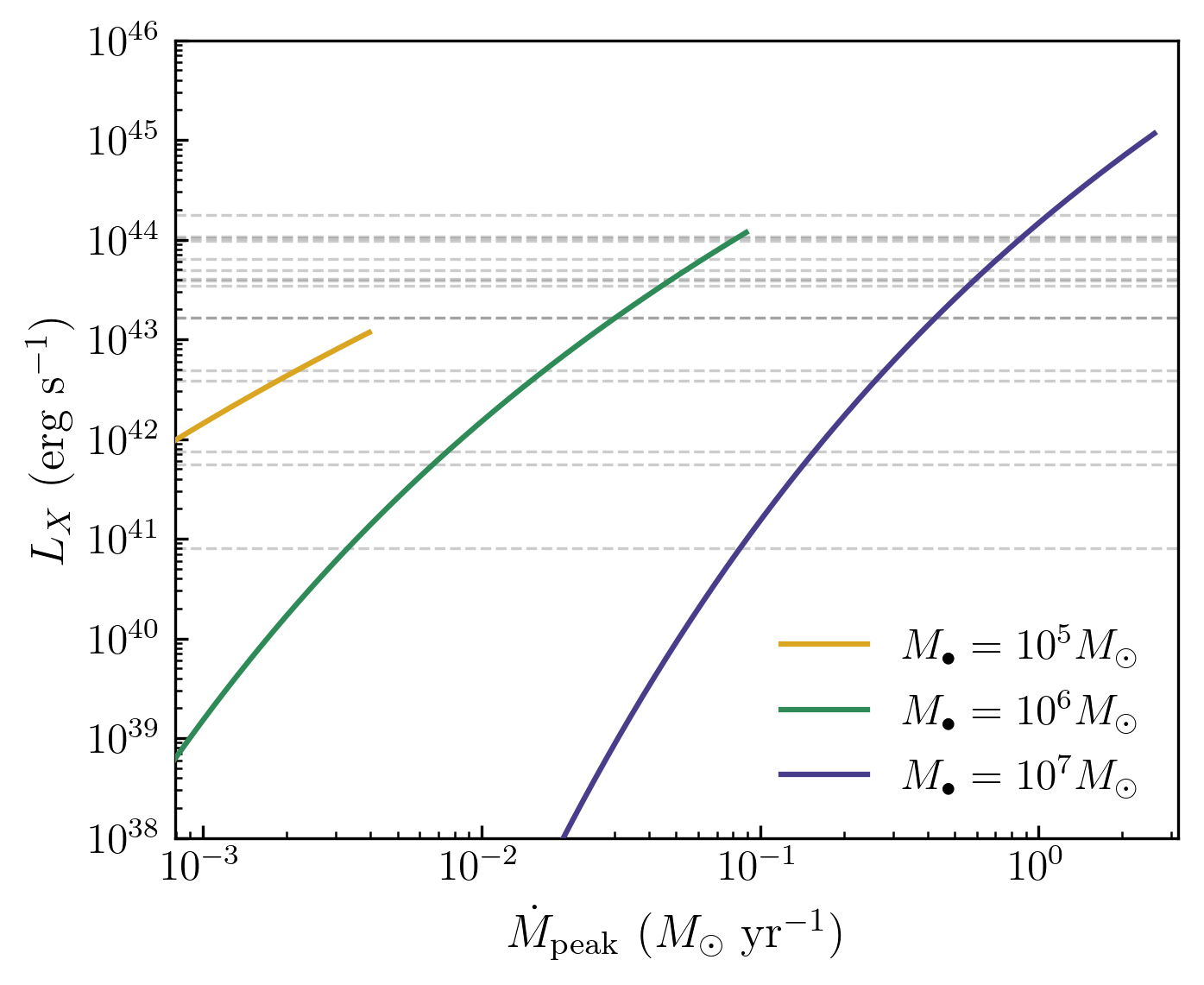}
\caption{Peak thermal disk X-ray luminosities as a function of $\dot{M}_\mathrm{peak}$, for a few different BH masses. These correspond to X-ray emission   integrated in the 0.3-10 keV band (as in observations), and have been limited to the sub-Eddington regime. Gray dashed lines represent peak X-ray luminosities for 17 different X-ray detected TDEs, as presented in \cite{guolo2024}.  The inferred accretion rates to power the observed thermal X-ray emission span a factor of $\sim 10-100$, plausibly corresponding to a range of disk viscous times, from disks that spread significantly over several years to those that don't.}
\label{fig:Lx_Mdot}
\end{figure}

The results we obtain through these estimates are consistent with our full radial, time-dependent simulations. We present the analytic estimates in Figure \ref{fig:Lx_Mdot}, where we show the  peak X-ray luminosity $L_X$ in the 0.3-10keV band as a function of $\dot{M}_\mathrm{peak}$ for three different BH masses. We have included the data from \cite{guolo2024} as dashed lines for reference (their inferred BH masses are also consistent with those we consider in our models).   Taking a fiducial BH mass of $\sim 10^{6} M_\odot$, Figure \ref{fig:Lx_Mdot} implies $\dot M_{\rm peak} \sim 0.01- 0.1M_\odot \, {\rm yr^{-1}}$ to explain the observed X-ray data.    This in turn corresponds to initial disk viscous times $\sim 1-10$ yr given a fiducial initial disk mass of $M_{d,0} \sim 0.1 M_\odot$.  This is fully consistent with the range of conditions needed to explain the optical plateau luminosities in Figure \ref{fig:constant_lum}. The higher X-ray luminosity systems have shorter inferred viscous times and are likely undergoing significant spreading over the time of the late-time optical-UV observations, while the lower X-ray luminosity systems have longer inferred viscous times and are consistent with not spreading significantly on year timescales.  The exact quantitative constraints are, however, subject to the uncertainty in the initial disk mass $M_{d,0}$ and do not directly constrain the initial disk radius $R_{d,0}$ since that is degenerate with $\alpha (H/R)^2$ in setting $\dot M_{\rm peak}$ (eq. \ref{eq:Mdot_peak}). It is also important to point out that these estimates are approximate since they do not account for relativistic effects in the inner disk that can modify the X-ray spectrum.

Most soft X-rays seen in TDEs are quasi-thermal  \citep{guolo2024}. In our magnetically elevated disk models, the disk surface density at small radii decreases  compared to a simpler Shakura-Sunyaev  model, which raises the question of whether the X-ray producing radii are sufficiently optically thick to reach thermal equilibrium consistent with the observations.

In order to assess this, we derive an analytical scaling for the absorption optical depth at the last stable circular orbit (for a = 0) assuming free-free absorption

\begin{equation}
    \begin{split}
        \tau_\mathrm{abs}(r_\mathrm{ISCO})\approx0.03&\left(\frac{\alpha}{0.01}\right)^{1/2}\left(\frac{\dot{M}_\mathrm{peak}}{0.05M_\odot\text{yr}^{-1}}\right)^{1/4}\\
        &\left(\frac{M_\bullet}{10^6M_\odot}\right)^{-3/8}\left(\frac{\alpha(H/R)^2}{10^{-4}}\right)^{-13/8};
    \end{split}
    \label{eq:tau_abs}
\end{equation}
the corresponding scattering optical depth is

\begin{equation}
    \begin{split}
        \tau_\mathrm{sc}(r_\mathrm{ISCO})\approx10^5\left(\frac{\dot{M}_\mathrm{peak}}{0.05M_\odot\text{yr}^{-1}}\right)&\left(\frac{M_\bullet}{10^6M_\odot}\right)^{-1}\\
        &\left(\frac{\alpha(H/R)^2}{10^{-4}}\right)^{-1},
    \end{split}
    \label{eq:tau_sc}
\end{equation}
such that the effective optical depth is given by

\begin{equation}
\tau_\mathrm{eff}(r_\mathrm{ISCO})=\sqrt{\tau_\mathrm{abs}\left(\tau_\mathrm{abs}+\tau_\mathrm{sc}\right)}\approx50
\end{equation}
for our fiducial values used in equations (\ref{eq:tau_abs}) and (\ref{eq:tau_sc}). In fact, all parameter combinations plotted in Figure \ref{fig:Lx_Mdot} have $\tau_\mathrm{eff}>1$, such that they would be consistent with the observed quasi-thermal nature of the X-ray emission.

\subsubsection{Irradiation}
\label{subsec:results_irradiation}

\begin{figure}
\centering
\includegraphics[width=\linewidth]{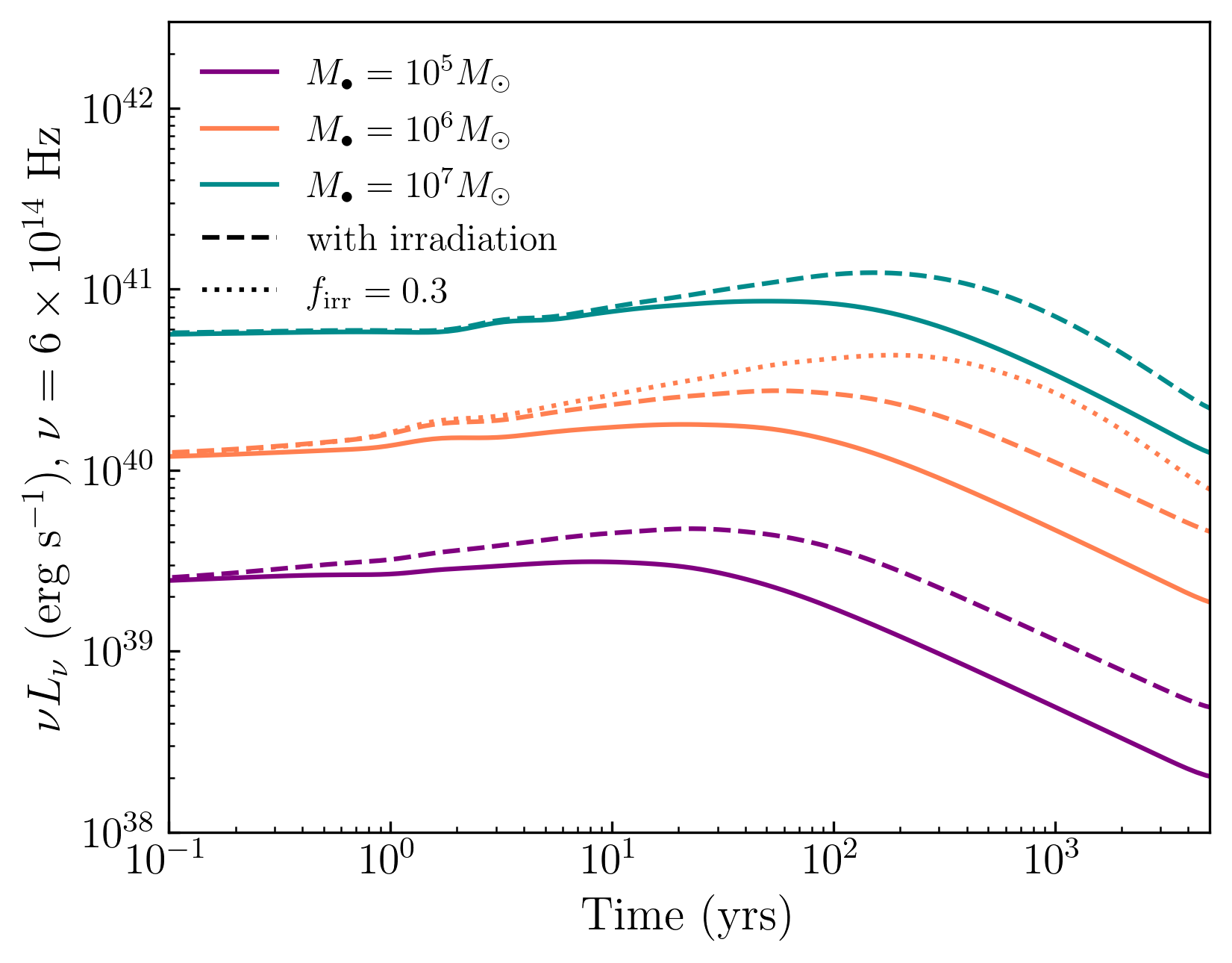}
\caption{Optical light curves of magnetized disks, with (dashed and dotted) and without (solid) irradiation of the outer warped disk by the inner disk. The dashed lines use $f_\mathrm{irr}=\sin\theta_\mathrm{irr}$ (see \S\ref{subsec:irradiation}), while the dotted line takes a constant $f_\mathrm{irr}=0.3$ as a likely upper limit. The initial disk mass is calibrated by mass loss due to super-Eddington fallback, set by equation (\ref{eq:Md0}). We consider fiducial values of $\alpha=0.01$ and $\theta_\star=\pi/6$ for these models. Irradiation from the inner disk boosts the late-time optical/UV luminosity by a factor of $\lesssim 2$, and further extends the plateau phase.}
\label{fig:irradiation}
\end{figure}

We show the effects of irradiation on the emission from the spreading disks in TDEs in Figure \ref{fig:irradiation}. The additional heating from irradiation, which becomes comparable to the viscous heating at large radii, increases the temperature of the outer disk and increases its luminosity in the optical/UV. We see some brightening compared to the non-irradiated models after a few years, which also extends the luminosity plateau to somewhat later times.  The dotted line in Figure \ref{fig:irradiation} shows a likely upper limit on the effects of irradiation, in which we assume that a fixed fraction $f_\mathrm{irr} = 0.3$ of the inner disk luminosity irradiates the outer disk.  This is likely only realizable if the outer disk is heavily flared and/or outflows scatter much of the inner disk radiation down onto the outer disk.

\subsection{LFBOTs}

\begin{figure}
\centering
\includegraphics[width=\linewidth]{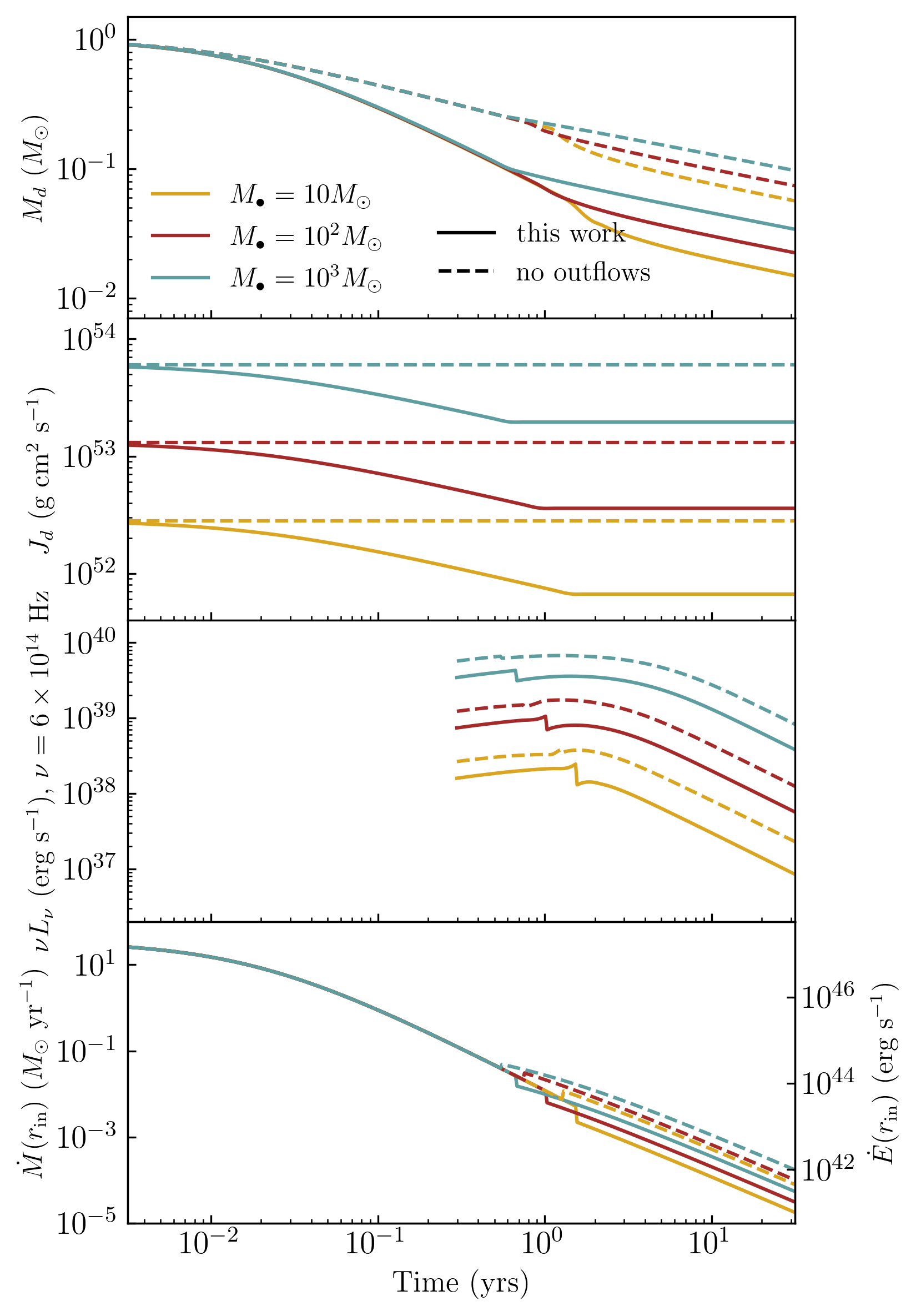}
\caption{Temporal evolution of magnetized disks around a $10$, $100$ and $10^3M_\odot$ BH using the one-zone model, for initial conditions appropriate for a BH-star merger  (i.e., a bound star not a parabolic TDE). The initial disk mass is set to $M_{d,0}=1M_\odot$ and we take $\alpha=0.01$ for all models. Solid lines account for mass loss due to outflows when the outer edge of the disk is super-Eddington. Dashed lines represent the standard model without outflows, such that the angular momentum of the disk is conserved. Early-time luminosities are not plotted as the thin disk approximation breaks down when emission is super-Eddington. Models including outflows show a more dramatic super-Eddington to sub-Eddington transition in their luminosity curves, and predict lower late-time luminosities than the standard picture without outflows.}
\label{fig:fbot_evolution}
\end{figure}
LFBOT emission across the electromagnetic spectrum (radio-X-ray) \citep{Margutti2019}, and their off-nuclear locations in star-forming galaxies \citep{Ho2020},
represent a new class of extragalactic transients.   Their origin remains uncertain but the need to explain luminous radio, optical, and X-ray emission has led to models centered on either ``failed'' collapses of massive stars to BHs \citep{Chrimes2025}, stellar mass BH mergers with stars in binary systems \citep{Metzger2022}, or the tidal disruption of stars by stellar (10--100\,$M_\odot$; \citealt{Kremer2023}) or intermediate ($10^{3}$--$10^{4}\,M_\odot$; \citealt{Perley2019,kuin2019}) mass BHs.   The evidence for a similar circumstellar medium environment in several LFBOTs \citep{Nayana2025}  is suggestive of pre-merger mass transfer of stars in binary systems \citep{Klencki2025}.   Intriguingly, the prototypical LFBOT AT2018cow shows late-time emission in the optical-UV (and possibly X-ray) $\sim 5$ years after the event \citep{sun2022,Chen2023, Migliori2024,inkenhaag2023,inkenhaag2025}.  This optical-UV detection and the lack of bright X-rays has been used to argue against stellar mass BH models for LFBOTs because an extrapolation of thin disk spectral models to the X-rays significantly overproduces the X-ray emission \citep{Chen2023}. 

The slow fading in the optical-UV at late-times in AT2018cow bears some similarity to the optical-UV plateau in TDEs.  In this section we explore this analogy by constructing disk models for the late-time emission in LFBOTs.  We specifically focus on the case of bound stars on circular orbits, i.e., stellar mergers.   We suspect that our results are qualitatively relevant to TDE models but as discussed in \S \ref{subsec:superEdd fallback}, the fallback rate for TDEs is super-Eddington for decades for stellar and intermediate mass BHs so the physics of disk formation is particularly uncertain in this regime.   In our models in this section we use the one-zone model to estimate the evolution of the overall disk properties incorporating outflows from the super-Eddington outer edge of the disk.  Given the disk mass and radius as a function of time we can further estimate the optical emission (which will be dominated by the outer portion of the disk) and the accretion rate close to the central BH, which will set the X-ray luminosity and the total outflow power.

In Figure \ref{fig:fbot_evolution} we show our results for the time evolution of the disk mass $M_d$, angular momentum $J_d$, optical luminosity, and accretion rate, for different BH masses. To calculate the luminosity, as we do not have the full radial temperature profile due to the one-zone model framework, we assume the temperature to be that of a steady state disk with our estimated accretion rates. We then integrate this blackbody up to the outer radius of the disk to obtain the spectral luminosity. In this plot we also compare to models that do not account for outflows.   The sharp changes in $M_d(t)$, $J_d(t)$, and $\nu L_\nu(t)$ at around 1 year represent when the outer disk becomes sub-Eddington, and the evolution transitions to angular momentum conserving (the discontinuity in the luminosity curves reflects the change from $H/R = 1/3$  in the super-Eddington phase to an $H/R$ set by magnetic fields as the accretion rate drops below Eddington).\footnote{If radiation dominated thin disks were stable, the evolution from super to sub-Eddington would be continuous.   The models here assume that radiation dominated thin disks are unstable, which leads to a discontinuous transition in disk properties from the super-Eddington phase to the magnetically supported thin-disk phase.}   We do not plot the luminosities before $t\sim0.3$yr in Figure \ref{fig:fbot_evolution}, as at these early times even the outer disk is still super-Eddington and therefore the thin disk approximation we use to calculate the luminosity is not valid. 
Although not the focus of this paper, we note that the high speed outflows from super-Eddington accretion  have broadly the right properties to explain the circumstellar interaction powered radio emission in AT2018cow \citep{ho2018}, as emphasized in \citet{Metzger2022}.   Most of the outflow energy is produced in the first few viscous times; the radio emission on longer timescales is analogous to that produced by a point explosion, making it difficult to distinguish between wind-powered and explosion powered radio emission.

From Figure \ref{fig:fbot_evolution} we can see that the optical luminosity plateaus for some time when the outer disk transitions from super-Eddington to sub-Eddington, before starting to decay at later times. This plateau phase is brief for the $M_\bullet=10M_\odot$ case, but becomes more extended as we go up in BH mass. Given our model for the evolution of the super-Eddington accretion phase (\S \ref{subsubsec:superedd outflows}), the time at which the outer disk accretion transitions to sub-Eddington is given by 
\begin{equation}
\begin{split}
    t_\mathrm{Edd,out}\approx 1.5\text{ yr}& \left(\frac{M_\bullet}{100M_\odot}\right)^{-1/7}\left(\frac{\alpha}{0.01}\right)^{-4/7}\\
    &\left(\frac{M_\star}{10M_\odot}\right)^{2/7}\left(\frac{R_\star}{0.8R_\odot}\right)^{3/7},
    \label{eq:t_edd,out}
    \end{split}
\end{equation}
where we have scaled to stellar properties appropriate for Wolf-Rayet stars (motivated by \citealt{Metzger2022}).   Equation (\ref{eq:t_edd,out}) can then be used to estimate $M_d(t_\mathrm{Edd,out})$, $R_d(t_\mathrm{Edd,out})$,  and the effective temperature at the outer edge of the disk, which is where most of the UV/optical emission comes from,
\begin{equation}
\begin{split}
    T_\mathrm{eff}(t_\mathrm{Edd,out})\sim 4&\times10^4\text{ K}\left(\frac{M_\bullet}{100M_\odot}\right)^{4/147}\left(\frac{\alpha}{0.01}\right)^{-15/98}\\
    &\left(\frac{M_\star}{10M_\odot}\right)^{-23/294}\left(\frac{R_\star}{0.8R_\odot}\right)^{-23/196}.
\end{split}
\end{equation}
For cases in which the outer disk emission lies on the Rayleigh-Jeans tail, we can make a simple estimate of how the luminosity plateau scales with BH and stellar parameters, analogous to (\ref{eq:Lscalings}) obtained for the TDE case: 
\begin{equation}
\begin{split}
    \nu L_\nu\sim 3 \times&10^{39}\text{ erg s$^{-1}$}\left(\frac{\nu}{6\times10^{14}\text{ Hz}}\right)^3\left(\frac{M_\bullet}{100M_\odot}\right)^{74/147}\\
    &\left(\frac{\alpha}{0.01}\right)^{41/98}
    \left(\frac{M_\star}{10M_\odot}\right)^{89/294}\left(\frac{R_\star}{0.8R_\odot}\right)^{89/196},
    \end{split}
    \label{eq:LRJFBOT}
\end{equation}
\begin{figure}
\centering
\includegraphics[width=\linewidth]{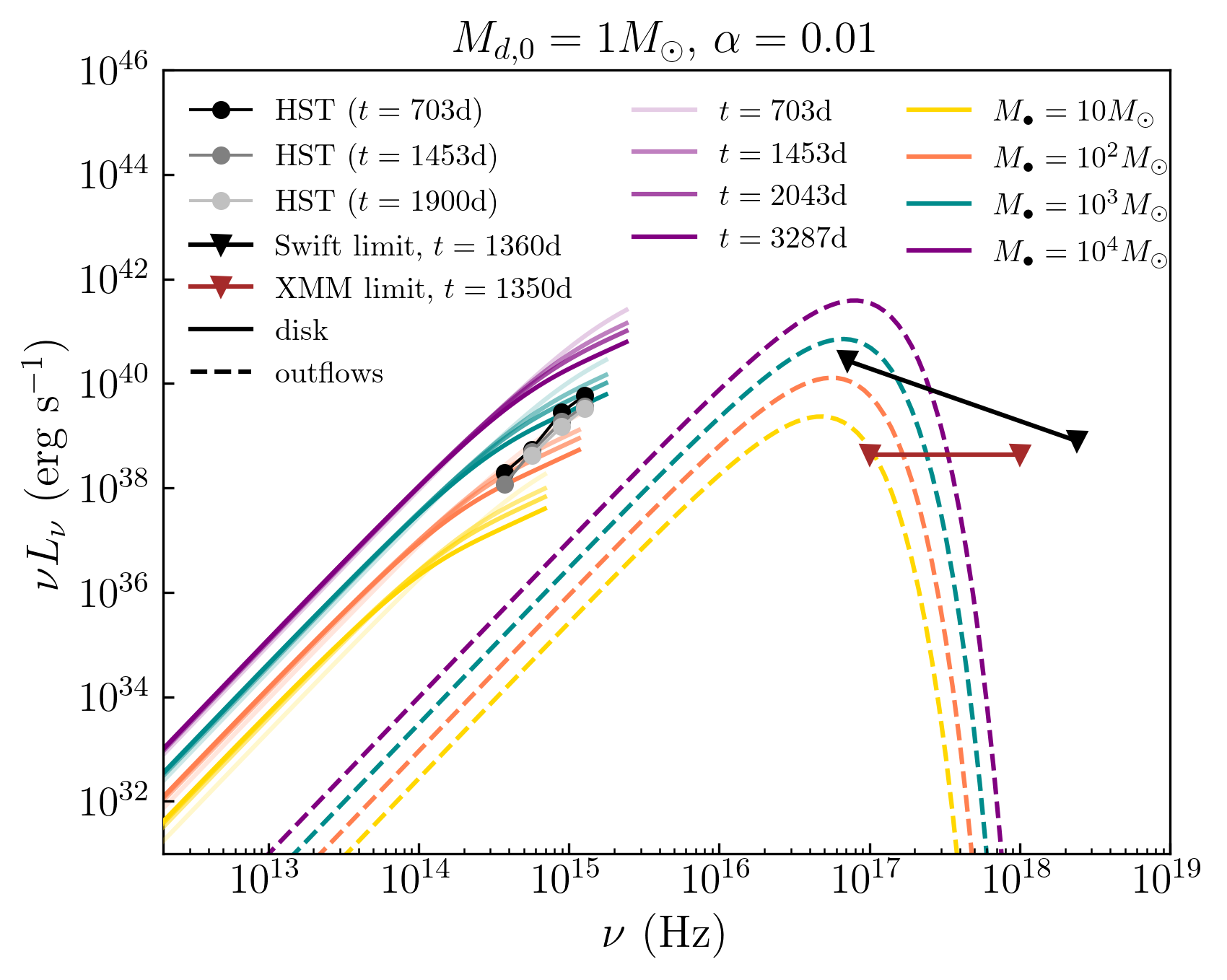}
\caption{SEDs for the outer thin disk  (solid) and super-Eddington outflows from the inner disk (dashed) for different BH masses; these results use initial conditions appropriate for a BH-star merger (i.e., a bound star not a parabolic TDE). The solid lines, from more transparent to more opaque, correspond to $t=703$d, $t=1453$d, $t=2043$d and $t=3287$d respectively. The thin disk SEDs are truncated at $h\nu=3k_BT_\mathrm{eff}(r_\mathrm{Edd})$ as higher frequencies correspond to radii where the emission becomes super-Eddington, and therefore the thin disk approximation is no longer valid. HST data is shown for three different epochs, however we note that there could be an uncertain starlight contribution to the emission. The outflows for the SEDs are plotted at a time $t=1360$d consistent with the $3\sigma$ Swift-XRT upper limits on AT2018cow (downward triangles) which assume a power law with a fiducial photon index of $\Gamma=3$. XMM detections at $t=1350$d are also included in dark red. Swift is much less sensitive than XMM in the soft X-ray, which is reflected by the plotted upper limits. For the outflow spectra we consider a fiducial inner disk wind speed of  $v_w = 0.3$c. Models with $M_\bullet\gtrsim100M_\odot$ are incompatible with the AT2018cow observations as they are too bright bolometrically.  See \S \ref{sec:Xrayabsorption} for a discussion of the role of X-ray reprocessing into the optical-UV at late-times.}

\label{fig:FBOT_BH_SEDs}
\end{figure}

The approximation in equation (\ref{eq:LRJFBOT}) is more applicable for higher mass BHs and/or higher disk masses, as for lower masses the emission from the outer disk starts falling off the Rayleigh-Jeans tail (see Figure \ref{fig:FBOT_BH_SEDs} for a range of models).   A comparison to the late-time data in AT2018cow also shown in Figure \ref{fig:FBOT_BH_SEDs} suggests BH masses of perhaps $\sim 10^{2-3} M_\odot$; we will now argue that somewhat lower BH masses of $10^{1-2} M_\odot$ are favored when accounting for the late-time X-ray limits and allowing for disk masses with $M_{d,0} > 1 M_\odot$ (indicative of more massive companions in stellar merger models).

When the outer disk transitions to locally sub-Eddington accretion on a timescale of $t_{\rm Edd,out}$ given by equation (\ref{eq:t_edd,out}), the inner accretion flow remains highly super-Eddington.   Thus the emission from the inner disk in LFBOT models with stellar mass BHs cannot be predicted using thin disk models, as has been done previously in the literature (e.g., \citealt{Chen2023}).    Instead, we use the models of \citet{strubbe_optical_2009, linial_tidal_2024} to estimate the radiation produced by the super-Eddington outflows from smaller radii. We focus on the outflow produced near the inner edge of the accretion flow because in the simplest super-Eddington wind models the radiated luminosity is $L \sim L_{\rm Edd}^{2/3} \dot E^{1/3}$ for a wind of total power $\dot E > L_{\rm Edd}$. The smallest radii dominate the wind power $\dot E$ and thus the total outflow luminosity.

We approximate the outflow as a quasi-spherical, radiation-dominated outflow of velocity $v_\mathrm{w}=\beta_\mathrm{w} c$, which carries away a significant fraction $f_\mathrm{w}\lesssim1$ of the total accreted mass at small radii $\dot{M}_\mathrm{w}=f_\mathrm{w}\dot{M}(r_\mathrm{in})$ (with $\dot M({r_{\rm in}})$ estimated from eq. \ref{eq:Mdot_in}). For simplifying purposes the following assumptions are made: (a) the outflow is steady-state and traces the accretion rate onto the BH, $\dot{M}_\mathrm{in}$; (b) the outflow is spherical and propagating at a fixed velocity $v_\mathrm{w}$ independent of $\dot{M}_\mathrm{in}$; and (c) the outflow is launched near the inner edge of the disk at roughly the sonic radius $r_s=r_g\beta_\mathrm{w}^{-2}$. Following equations (16) and (17) from \cite{linial_tidal_2024} for the outflow's bolometric luminosity and blackbody temperature, we can estimate the spectral luminosity produced by the outflow as
\begin{equation}
    L_\nu\approx\frac{\pi B_\nu(T_\mathrm{BB})L}{\sigma T_\mathrm{BB}^4},
\end{equation}
where the bolometric luminosity $L\propto\beta_{\rm w}^{2/3}$, $T_{\rm BB}^4\propto \beta_{\rm w}^{-1/3}$ and we take $\beta_\mathrm{w} \sim 1/3$ as these outflows are launched from the vicinity of the BH. Although these scalings are intentionally simplified, recent radiation-hydrodynamic simulations actually find that super-Eddington TDE flows naturally generate quasi-spherical radiation-driven winds with adiabatic losses and Eddington-regulated emission broadly consistent with the analytic scalings adopted here \citep[e.g.][]{martire2025}.

We show how the emission coming from both the outer edge of the thin disk and the outflow at the inner disk vary with BH mass in Figure \ref{fig:FBOT_BH_SEDs}, where data points for AT2018cow observations have been included for reference. We plot the thin disk SEDs only up to frequencies of $h\nu=3k_BT_\mathrm{eff}(r_\mathrm{Edd})$, as higher frequencies correspond to radii where the emission becomes super-Eddington, and therefore the thin disk approximation is no longer valid.  In reality the spectrum in between the optical and X-ray will be filled in by outflows produced exterior to $r_{\rm in}$ which we are not including in the simple models developed here.    The dashed curves in Figure \ref{fig:FBOT_BH_SEDs} representing the outflow emission are dominant in the soft X-ray.  Even allowing for the possibility that the X-ray emission is reprocessed by gas at larger radii into the optical-UV (as discussed below), models with $M_\bullet\gtrsim10^2M_\odot$ are incompatible with the AT2018cow observations as they are too bright both in the optical/UV and in the soft X-rays. 

\begin{figure}
\centering
\includegraphics[width=\linewidth]{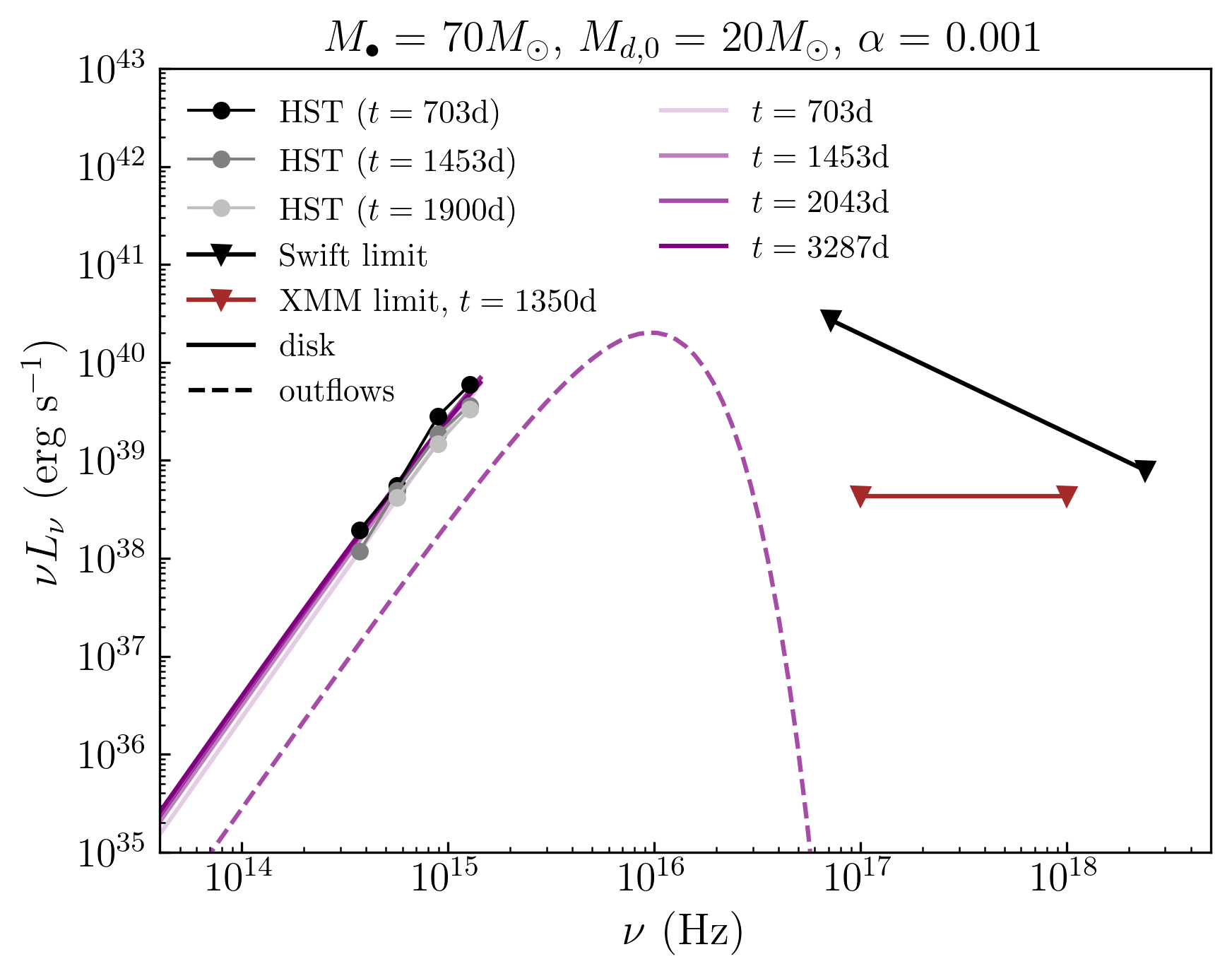}
\caption{Example SED (analogous to those in Figure \ref{fig:FBOT_BH_SEDs}) for a combination of $M_\bullet$, $M_{d,0}$ and $\alpha$ that provides a model consistent with AT2018cow observations for stellar-mass BH models.   The far UV-X-ray predictions are more uncertain and are plausibly soft enough to be compatible with the upper limits.   The outflow emission is softer here than in Figure \ref{fig:FBOT_BH_SEDs} because the more massive disk and lower $\alpha$ leads to higher outflow rates with a cooler photosphere.}
\label{fig:FBOT_examples}
\end{figure}

We show an example of a model that can broadly explain the late-time data in AT2018cow in Figure \ref{fig:FBOT_examples} (multiple combinations of $M_\bullet$, $M_{d,0}=M_\star$ and $\alpha$ can do so, so this is just illustrative). We consider our modeling of the thin disk emission to be quite reliable, while there is much more uncertainty related to the outflow emission.  In this particular model, the outflow emission is soft enough to produce negligible X-rays that would be detectable by Swift and/or XMM.   This is because the larger disk mass and lower $\alpha$ in this model lead to a higher late-time mass outflow rate which produces softer outflow emission ($T_{\rm BB} \propto \dot M_w^{-5/12}$).

\begin{figure}
\centering
\includegraphics[width=\linewidth]{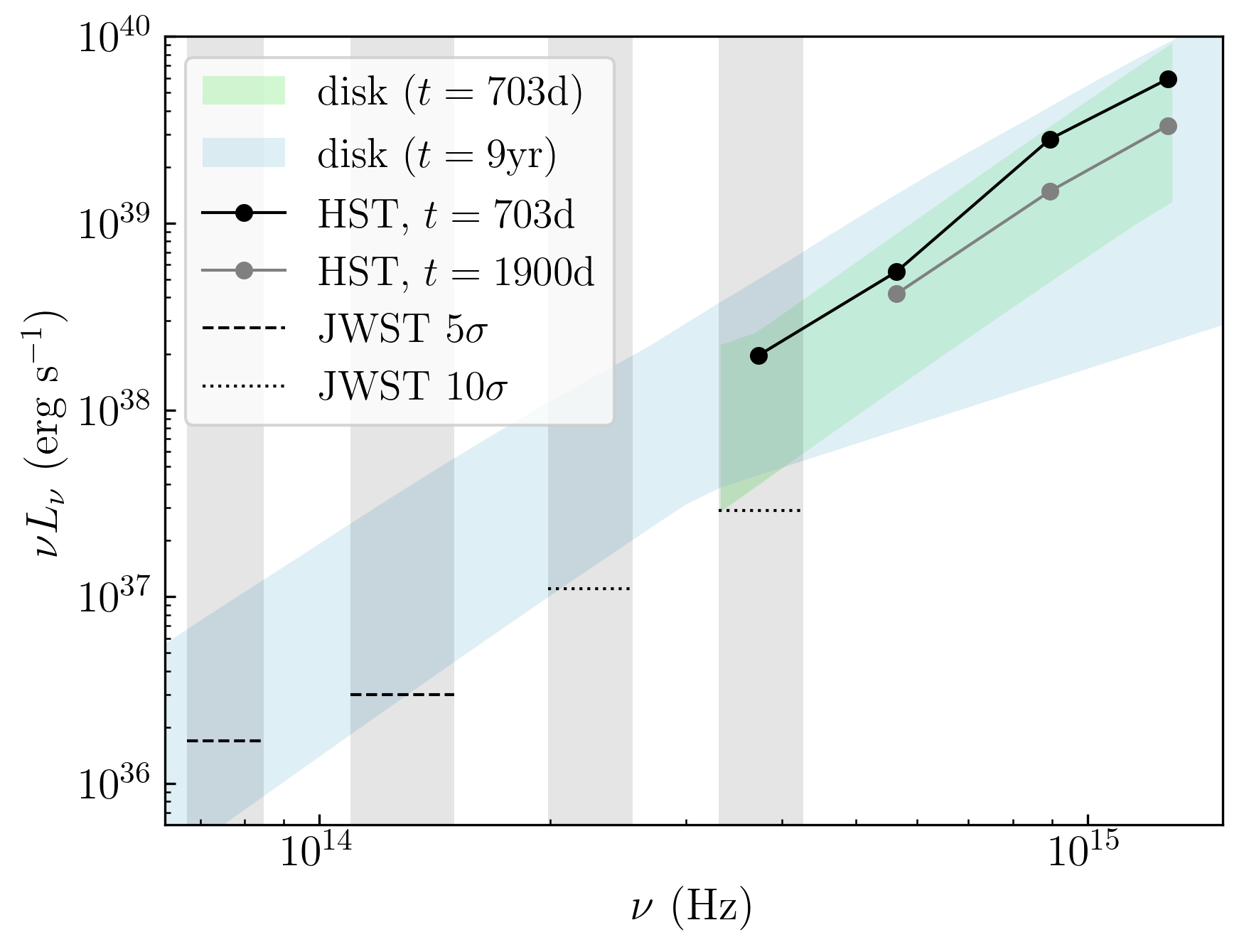}
\caption{Prediction for $t=9$ years of models consistent with the HST data for AT2018cow. The green shaded region shows the range of luminosities covered by models that are within a factor of few of the $t=703$d observations, and the blue shaded region shows predictions of what these SEDs will look like at $t=9$ years. The gray vertical bands indicate from left to right the F356W, F277W, F150W and F070W JWST bands along with their estimated sensitivities, as a reference for potential future observations. The fainter predicted SEDs are produced by the less massive BHs, and present a RJ break that moves through the optical. Observing this behavior would thus be a strong confirmation of a low mass BH model.}
\label{fig:predictions}
\end{figure}

We expand on the range of late-time models by showing a large range of predictions for the UV/optical emission of AT2018cow using the thin disk emission in Figure \ref{fig:predictions}. These are obtained by generating SEDs for all combinations of $M_\bullet=10-10^4M_\odot$, $M_\star=0.5-20M_\odot$ and $\alpha=10^{-3}-1$, and selecting those that are consistent with the data at $t=703$d within a factor of a few. For these purposes we take the HST detections as upper limits, considering that there may be an underlying host galaxy contribution \citep{inkenhaag2023,inkenhaag2025}. The range covered by these models is shown by the region shaded in green in Figure \ref{fig:predictions}. We then obtain the SEDs for these same models at $t=9$ years, and plot their range in blue. These are our predictions of what should be observed by $t \sim 9$ yr.   We note that the lower BH masses take up the lower bounds of the predicted SEDs, suggesting that if we were able to observe the RJ break move through the optical it would be a strong confirmation of a low mass BH. We also include in this figure the sensitivities for a few JWST bands to show potential for future observations to further constrain the late-time emission in AT2018cow, which could bring us one step closer to understanding the origin of LFBOTs.

\subsubsection{X-ray Absorption}
\label{sec:Xrayabsorption}

A surprising feature of the late-time emission in AT2018cow is that the X-ray luminosity is lower than the optical luminosity.   Although both may have some host galaxy contribution the optical emission has faded in time indicating a significant contribution from the transient.   The optical to X-ray ratio is not easily explained by accretion models as indicated by Figure \ref{fig:FBOT_BH_SEDs}.    The basic reason is that the inner X-ray emitting region is generically more luminous than the outer optical emitting region in models with stellar or intermediate mass BHs.  This effect is less prominent in our models than in previous work (e.g., \citealt{Chen2023} Fig. 10) because we account for the transition from thin disk emission at large radii to super-Eddington accretion and outflows at smaller radii.   

This tension is alleviated in the model in Figure \ref{fig:FBOT_examples} by having the super-Eddington outflow emission from small radii peak in the far UV.  Given the uncertainty in the outflow emission it is possible that a model with roughly these properties  is compatible with the observations.    Another possibility, however, is that at late-times the X-ray emission produced by the inner outflow is reprocessed to longer wavelengths by gas at larger radii.  Qualitatively, this would again require stellar mass BHs $\lesssim 100 M_\odot$ so that the X-ray emission from the outflows in Figures \ref{fig:FBOT_BH_SEDs} and \ref{fig:FBOT_examples} is comparable to or below the HST optical detections.

We explore the possibility of X-ray outflows coming from the vicinity of the BH being absorbed by gas produced by the outer outflows described in section \ref{subsubsec:superedd outflows}, and reprocessed into UV/optical emission.    The efficiency of this absorption/reprocessing depends on the ionization parameter of the surrounding gas irradiated by the X-rays
\begin{equation}
\xi = \frac{1}{4\pi r^2 n c}\int_{\varepsilon_0}^\infty\frac{L_\nu}{h\nu} \, d\nu,
\label{xi}
\end{equation}
with $\varepsilon_0=1$keV the oxygen ionization threshold if we are interested in the absorption above a keV or so.   We replace the number density in equation (\ref{xi}) in favor of an outflow rate using

\begin{equation}
\label{eq:n(r)}
    n(r)\approx\frac{\dot{M}_\mathrm{out}}{f_\Omega4\pi r^2 \mu m_p v_\mathrm{out}},
\end{equation}
where $\dot{M}_\mathrm{out}$ is as defined in equation (\ref{eq:Mdot_out}), $v_\mathrm{out}\sim\sqrt{GM_\bullet/r_\mathrm{out}}\ll c$ is the velocity of the outflow originating primarily at radius $r_{\rm out}$, $\mu=0.6$ is the mean molecular weight and $f_\Omega$ is the fraction of solid angle that the outflow covers. The ionization parameter can then be written as $\xi \sim f_\Omega L_x \mu m_p/(\dot M_{\rm out}\varepsilon_0)(v_{\rm out}/c)$.  Using $L_X \sim \dot E_{\rm in}^{1/3} L_{\rm Edd}^{2/3}$ for super-Eddington winds where $\dot E_\mathrm{in} \sim \dot M_\mathrm{in} \beta_{\rm w}^2 c^2$ and $\dot M \propto r^{1/2}$ (i.e., p = 1/2 per the discussion in \S \ref{subsubsec:superedd outflows}) we  find
\begin{equation}
\xi \sim f_\Omega \beta_{\rm w}^{2/3} \ \frac{\mu m_p c^2}{\varepsilon_0} \left(\frac{r_\mathrm{in}}{r_\mathrm{Edd}}\right)^{4/3}
\label{eq:xi2}
\end{equation}
where $r_\mathrm{in} \sim GM_\bullet/c^2$ and we have assumed that at late-times the absorbing outflows originate from the radius where the disk transitions to sub-Eddington, $r_\mathrm{Edd}\sim GM_\bullet \dot{M}/L_\mathrm{Edd}$.   This is conservative in that outflows produced at earlier times may also contribute some absorption.

\cite{govreen-segal_x-ray_2025} estimated a critical value $\xi_c=0.015$ that allows us to determine whether or not the X-rays efficiently ionize the ambient gas; for $\xi\ll\xi_c$ the gas is neutral and absorbs the x-rays, while for $\xi\gg\xi_c$ essentially all of the gas is ionized and therefore there is no X-ray absorption.

We obtain $\xi \ll1$ for our $M_\bullet=10M_\odot$ and $M_\bullet=100M_\odot$ models presented in Figure \ref{fig:FBOT_BH_SEDs}, evaluating the outflows at $t=1360$d to be consistent with the data we have for the soft X-ray emission for AT2018cow.  These estimates suggest that significant X-ray absorption by the late-time outflows is plausible. To further support this, we performed  calculations with {\fontfamily{cmtt}\selectfont Cloudy} for the $M_\bullet=100M_\odot$ model shown in Figure \ref{fig:FBOT_BH_SEDs} and the model shown in Figure \ref{fig:FBOT_examples}. We input the plotted X-ray SED as the incident emission, which interacts with an absorbing medium whose  density is given by eq. (\ref{eq:n(r)}). We find that the X-rays are efficiently absorbed for a range of mass outflow rates and velocities.  The emergent spectrum is not shown here because our models have high electron scattering optical depths for which {\fontfamily{cmtt}\selectfont Cloudy} is not very accurate.

This analysis motivates a modified scenario for the origin of the late-time emission in AT2018cow and perhaps other LFBOTs. Rather than the UV/optical emission coming only from the disk, what we are observing can be a combination of a disk and X-rays that have been reprocessed into UV/optical by outflows coming from the large radii in said disk. This would explain the fact that $L_{UV}\gtrsim L_X$ and help suppress the X-ray emission that would otherwise be expected.  This model would continue to favor lower mass BH models $\sim 10-100 M_\odot$ as such models produce lower X-ray luminosities (Fig. \ref{fig:FBOT_BH_SEDs}) broadly consistent with the optical/UV emission in AT2018cow if the latter has a significant contribution from reprocessing at late-times. 

Our goal in this section is only to motivate the plausibility of the late-time X-ray reprocessing scenario; more work is clearly needed to quantitatively flesh it out and to understand whether the time evolution of the optical and X-rays in LFBOTs can be self-consistently understood.  The scenario proposed here predicts that at sufficiently late-times, as $r_{\rm Edd}$ decreases, the ionization parameter will increase (eq. \ref{eq:xi2}), and eventually the X-rays from the central super-Eddington outflows and disk should be detectable with a luminosity set primarily by the BH mass.  During the phase where the inner disk remains super-Eddington we find that the ionization parameter increases roughly $\propto t^{3/2}$ while the X-ray luminosity decays slowly in time $\propto t^{-3/7}$ and gets harder over time.   Unfortunately the modest power-law increase in the ionization parameter likely means that the central X-ray source will only become unobscured decades or centuries after the original event (assuming that X-rays are currently absorbed by the outflow).

\section{Summary and discussion}
\label{sec:discussion}
In this work we have developed a model for the time-dependent evolution of viscously spreading accretion disks around BHs, motivated by the application to initially compact accretion disks formed in TDEs, stellar mergers with BHs, neutron star-BH mergers, and related events.  This generalizes previous work (e.g., \citealt{cannizzo, shen_evolution_2014, mummery_fundamental_2023}) by assessing the role of effects that are likely to be important in such systems:   mass loss and angular momentum redistribution in TDE circularization, outflows from super-Eddington accretion flows, irradiation of the outer warped disk by the inner accretion flow, and a range of viscous stress prescriptions.

The inclusion of outflows due to super-Eddington phases in our models has different implications depending on how the disk  forms. For a star on a parabolic orbit disrupted by a BH (i.e., TDEs), super-Eddington fallback likely suppresses the initial mass in the disk and leads to redistribution of angular momentum during circularization (e.g., \citealt{lu_bonnerot2020}) changing the initial angular momentum in the disk.  In stellar mergers all of the
star's mass and angular momentum initially ends up in the disk, but over time much of this mass and angular momentum is lost due to outflows driven by super-Eddington accretion.   The outflows subside at large radii once the outer disk becomes locally sub-Eddington, which occurs on a timescale of $\sim $ years for stellar mergers with stellar mass BHs (eq. \ref{eq:t_edd,out}).   Coincidentally, the fallback in parabolic TDEs by $\sim 10^{6-7} M_\odot$ BHs also becomes sub-Eddington on year timescales (eq. \ref{eq:tedd_fb}), which is likely when a thin disk forms.

We apply our models to the late-time optical-UV plateaus \citep{van_velzen_late-time_2019} observed in TDEs, which have been compellingly interpreted as signatures of thin disk formation and viscous spreading \citep{mummery_fundamental_2023}, as predicted by theoretical models once the initial super-Eddington fallback phase subsides \citep{cannizzo}.   

Current work has emphasized the role of viscous spreading to reproduce the late-time optical/UV luminosity plateaus and the fact that the disk luminosity on the Rayleigh-Jeans tail is only a weak function of time as the disk spreads \citep{mummery_fundamental_2023}.  We concur that this is compatible with the data.   We also find, however, that many of the late-time plateaus can be  qualitatively explained by disks with a large initial spread in angular momentum, i.e., with disk mass initially at radii exterior to that given by the specific angular momentum of the star (Fig. \ref{fig:constant_lum}).  The larger initial radius implies a significantly longer viscous time for the onset of spreading.  In addition, the luminosity on the Rayleigh-Jeans tail is only a weak function of viscous time $\propto t_{\rm{visc}}^{-1/4}$ (eq. \ref{eq:Lnospread}) which makes it not that sensitive to whether or not TDEs disks are in fact spreading significantly on the year-decade timescales observed thus far.  \cite{wen2023} similarly concluded that in the TDE ASASSN-14li, late-time observations were consistent with little to no disk spreading.   
The evolution of the early time SED in the far UV-X-ray and the peak X-ray luminosity is an independent probe of the viscous time of the bulk of the disk mass.    Shorter viscous times lead to higher peak accretion rates, higher X-ray luminosities, and more significant early-time far UV-X-ray evolution (Figs. \ref{fig:sigmas_seds} \& \ref{fig:Lx_Mdot}).   The observed peak X-ray luminosities of optically selected TDEs span a large range (e.g., \citealt{guolo2024}), perhaps suggesting a range of disk spreading timescales.

On timescales set by the viscous time of the outer disk formed during circularization, TDE disks will inevitably begin to viscously spread.  The timescale for such spreading depends on the poorly understood physics of angular momentum transport.  The simplest theoretical models predict that TDE disks will be radiation pressure dominated and thus thermally and viscously unstable \citep{Piran1978}.  We argue that collapse of radiation pressure dominated disks to the stable gas pressure dominated branch is ruled out by TDE plateaus, which are far too bright to be explained by gas pressure dominated TDE disk models (e.g., Fig. \ref{fig:TDE_Lnu_comparisons}).   Instead, the most likely possibility is that disks become magnetically supported, rendering them thermally and viscously stable (e.g., \citealt{begelman_accretion_2007}). Unfortunately the correct model for angular momentum transport in strongly magnetized accretion disks is unknown and current simulations even find multiple saturation states depending on the initial magnetic field in the simulation \citep{Zhang2025}.  It is also plausible that the saturated magnetic field in TDE disks will depend on the magnetic field in the star from which it formed.   Continued observations and modeling of late-time data in TDE disks offers an unparalleled opportunity to constrain the uncertain physics of accretion disk angular momentum transport.

The disks formed in parabolic TDEs are misaligned with respect to the spin of the central BH.   At late-times during the thin disk phase, this likely leads to the outer disk being strongly warped with respect to the inner disk that is aligned by the Bardeen-Petterson effect \citep{bardeen_petterson_1975}. If a substantial disk misalignment survives into the late-time thin-disk phase, irradiation of the warped outer disk may be important and should be included when modelling the emission. We show that such irradiation can boost the disk's luminosity and extend the duration of the plateau phase by factors of several (Fig. \ref{fig:irradiation}). 

\citet{mummery_fundamental_2023} showed that the late-time emission from TDE disks was a strong function of BH mass and proposed that the plateau emission could be used to measure BH mass.  We also find that the TDE disk emission is a strong function of BH mass (e.g., Fig. \ref{fig:constant_lum}-\ref{fig:irradiation}).  The late-time emission also depends on poorly understood aspects of accretion disk physics, including the physics of disk warps and irradiation, the dependence of angular momentum transport on disk properties (e.g., how $H/R$ and possibly $\alpha$ vary with disk properties), and how much of the star's mass and angular momentum end up in the late-time disk.  The exact scaling of late-time disk emission in TDEs with BH mass depends on any systematic variation of these effects with BH mass (eg. eq. \ref{eq:Lnospread} and eq. \ref{eq:Lscalings}).  Continued observational and theoretical study of late-time TDE emission has great potential to constrain both the host BH mass and these important aspects of accretion disk and TDE physics.

The second application we consider in this paper is to accretion disks formed in stellar mergers with stellar-mass BHs, motivated by the application to LFBOTs such as AT2018cow \citep{Margutti2019,ho+2023}, for which star-BH models are a possible model \citep{Metzger2022,Klencki2025}.   Our focus in this paper is solely on the late-time data in AT2018cow, and predictions for late-time emission in other systems (that is, we do not attempt to model the exquisite and remarkable early time observations of these systems!).   Years after the peak transient emission, AT2018cow shows a slowly fading optical-UV-X-ray source that bears some similarity to the late-time emission in TDEs \citep{Chen2023, Migliori2024}.   

Our primary conclusions in modeling AT2018cow's late-time emission are:  (1) Inclusion of super-Eddington outflows is critical in modeling the emission from star-BH mergers because such outflows can remove $\gtrsim 3/4$ of the disk's angular momentum and mass before the outer disk becomes sub-Eddington (Fig. \ref{fig:fbot_evolution}). (2) If the late-time emission in AT2018cow indeed arises from an accretion disk, our models favor $M_\bullet\approx10-100M_\odot$ BHs disrupting stars of masses around $1-30M_\odot$. These parameter combinations ensure that the optical/UV luminosity is bright enough to be consistent with data from HST, but that the X-ray emission coming from the inner outflows is faint and/or soft enough to  not be in too much tension with the current constraints from Swift and XMM. (3) Previous work highlighted the difficulty explaining both the late-time optical-UV and X-ray in AT2018cow with stellar mass BHs because extension of thin disk models from the optical-UV to X-ray dramatically overpredict the latter (e.g., Fig 10 of \citealt{Chen2023}).  Such models were inconsistent in assuming that the inner X-ray emission was from a thin disk, despite the very super-Eddington accretion rates needed in LFBOT models. We argue that the late-time X-rays may instead be from super-Eddington outflows. These models, although quite uncertain, may in some cases be soft enough  to satisfy the Swift and XMM upper limits in AT2018cow (Fig. \ref{fig:FBOT_BH_SEDs} and \ref{fig:FBOT_examples}).  We also suspect that the late-time far UV and X-ray emission is reprocessed into the optical-UV by ongoing outflows, because the X-ray luminosity is sufficiently low at late-times that the outflows cannot be fully ionized by the central X-ray source (\S \ref{sec:Xrayabsorption}). (4) Our models predict that late-time X-rays at $\sim 10^{39-40} \, {\rm erg \, s^{-1}}$ should eventually be detectable in LFBOTs once the ionization parameter of the surrounding outflows increases and that HST-JWST observations of AT2018cow may detect a break in the optical-IR spectrum, providing a powerful probe of the outer accretion disk thermodynamics (Fig. \ref{fig:predictions}).

The models of late-time LFBOT emission developed here could be improved in a number of ways.   Most importantly, we have considered a simple superposition of the emission from an inner super-Eddington outflow and an outer thin disk in our spectral models (e.g., Figs. \ref{fig:FBOT_BH_SEDs}).  A more careful treatment would require a better understanding of how outflows from different radii interact to produce the observed spectrum and how/whether the outer thin disk emission is itself reprocessed by the outflows.   We leave this to future work.   It is also unclear if our models can be applied to neutron star models of LFBOTs (e.g., \citealt{Tsuna2025}).  The stellar mass LFBOT models developed here all have highly super-Eddington accretion near the central compact object even decades after the initial disk formation.  The physics of such accretion would be very different for neutron stars.  We have also assumed throughout that during phases of super-Eddington BH accretion the inflow rate decreases with radius $\propto r^{1/2}$.   This assumption is motivated by current numerical work on non-radiative disk spreading \citep{guo_cyclic_2025}, which we think should be a good approximation to highly super-Eddington accretion.  There is some evidence for this in  accretion  simulations with radiation (e.g., \citealt{Hu2022,Zhang2025}), but it has yet to be explicitly demonstrated for the long term  spreading of disks relevant to this work.

\begin{acknowledgments}
We thank  Anna Ho, Brian Metzger, Clement Bonnerot, Wenbin Lu, and Ben Margalit for useful discussions, and the referee for insightful comments that improved the paper.   This work benefited from interactions supported by the Gordon and Betty Moore Foundation through grant GBMF5076 and through interactions at the Kavli Institute for Theoretical Physics, supported by NSF PHY-2309135.
\end{acknowledgments}

\bibliographystyle{aasjournal}

\bibliography{main}

\end{document}